\documentclass{article}

\usepackage{arxiv}
\usepackage{cite}
\usepackage{amsmath}
\usepackage[utf8]{inputenc} % allow utf-8 input
\usepackage[T1]{fontenc}    % use 8-bit T1 fonts
\usepackage{hyperref}       % hyperlinks
\usepackage{url}            % simple URL typesetting
\usepackage{booktabs}       % professional-quality tables
\usepackage{amsfonts}       % blackboard math symbols
\usepackage{nicefrac}       % compact symbols for 1/2, etc.
\usepackage{microtype}      % microtypography
\usepackage{lipsum}
\usepackage[pdftex]{graphicx} % Inclusão de gráficos
\usepackage{tabularx}
\usepackage{float}
\usepackage{multicol} 
\usepackage{multirow}
\usepackage{threeparttable}
\usepackage{float}
\usepackage{bbm}
\usepackage{tablefootnote}
\usepackage{soul}
\restylefloat{table}

\title{Skills to not fall behind in school}

\author{
  Felipe Maia Polo\thanks{https://felipemaiapolo.github.io/} \\
  Institute of Mathematics and Statistics\\
  University of São Paulo\\
  São Paulo, Brazil\\
  \texttt{felipemaiapolo@gmail.com} \\
}

\begin{document}
\maketitle

\begin{abstract}
Many recent studies emphasize how important the role of cognitive and social-emotional skills can be in determining people's quality of life. Although skills are of great importance in many aspects, in this paper we will focus our efforts to better understand the relationship between several types of skills with academic progress delay. Our dataset contains the same students in 2012 and 2017, and we consider that there was a academic progress delay for a specific student if he/she progressed less than expected in school grades. Our methodology primarily includes the use of a Bayesian logistic regression model and our results suggest that both cognitive and social-emotional skills may impact the conditional probability of falling behind in school, and the magnitude of the impact between the two types of skills can be comparable.\footnote{The code used in this paper can be found in \url{https://github.com/felipemaiapolo/paper_skills}.}
\end{abstract}

% keywords can be removed

\section{Introduction}

In a recent report by the World Economic Forum \cite{world2015new} there is a clear paradigm shift when it comes to skills needed by today's children, youth and adults. People are rethinking the skills considered fundamental to humans in recent decades, introducing a new skill set. The authors call this core skill set “21st Century Skills”: in total there are 16 skills splitted into three categories, (i) Foundational Literacies, (ii) Competencies, and (iii) Character Qualities. The first category includes skills that encompass more technical knowledge such as math, literacy and financial education. On the other hand, the second includes skills such as creativity, communication, and the third includes skills such as curiosity, leadership, and persistence. It is important to point out that in the report itself, the authors classify the 10 skills belonging to the last two categories as part of the set of Social and Emotional skills, i.e. skills related to the way people interact to the outside world and to themselves. Figure \ref{fig:skills} was extracted from \cite{world2015new} and exposes the 16 skills for the 21st century:

\begin{figure}[H] 
\centering 
\includegraphics[width=0.8\textwidth]{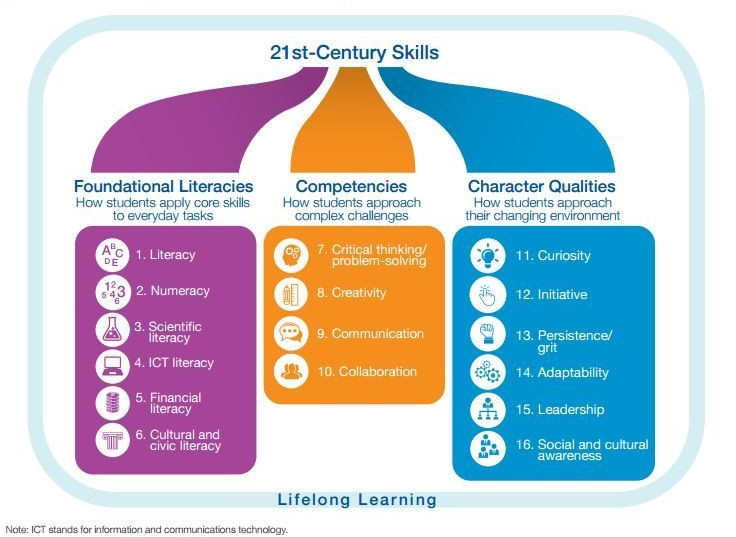} 
\label{fig:skills}
\caption[Caption for LOF]{The 16 Skills for the 21st Century \footnotemark}
\end{figure}
\footnotetext{Figure extracted from \citeonline{world2015new}.}

Just as we call the last 10 Social-Emotional skills we can call the first 6 as cognitive skills. Recent studies show findings by leading education researchers that these two types of skills take on great importance in defining people's future outcomes such as education, salaries and employment level \cite{heckman2006effects, kautz2014fostering}. Cognitive skills, for example, have a major impact on income and employability while Social-Emotional skills, which account for more than 60\% of fundamental 21st Century skills, have a positive impact on well-being and satisfaction and improving people's health through their lifestyle \cite{miyamoto2015skills}.

Although cognitive and social-emotional skills are of great importance in people's lives in many ways, in this paper we will focus on how these skills can affect the academic progress of students who were attending school in 2012 and 2017 in a city in the countryside state of São Paulo in Brazil. We have a rich dataset containing demographic, socioeconomic, and personal characteristics of students, and we can account for academic progress delay in a way that minimizes measurement errors - we consider that students who had academic progress delay between 2012 and 2017 were those who have progressed less in grades than expected, that is, those who fell behind in school. Our results suggest that, in fact, both cognitive and social-emotional skills impact how students progress in school.

\section{Literature}

In this paper we are concerned about the academic progress delays that can be caused either by school failure or temporary dropout. Failure and school failure can be explained from many angles, including personal skills, family and socioeconomic characteristics. Regarding failure, in addition to more direct variables such as cognitive skills, \cite{leon2002reprovaccao} gives evidence that variables such as income and parental education can have a large negative impact on the probability of failure. Regarding temporary dropout, some more concrete factors could be the need to enter the labor market prematurely \cite{arroyo1995educaccao}, lack of motivation and expectations about future studies return \cite{eckstein1999youths} or low parental education, which is a great proxy for the family's permanent income. In addition, students who have failed in the past form a group which is more vulnerable and prone to drop out \cite{soares2015fatores}. 

In a recent work \cite{carlos2019papel}, the author showed evidences that cognitive and social-emotional skills can have an impact on dropping out and reaching high school, which are variables of school progress. Although we use a dataset very similar to that work \cite{carlos2019papel}, we believe that this is an important paper to the literature because of the following factors: (i) In our work we used a larger number of social-emotional skills, which were not previously available and which are little explored in literature; (ii) Our dependent variable is a more general variable, which is academic progress delay, what helps us to deliver a broader message; (iii) We added an analysis of the odds of falling behind in school, which is natural within logistic regression framework; (iv) We use a Bayesian methodology, which among many benefits, we can highlight the fact that we will obtain the entire posterior distribution of the parameters of interest and not just a point estimate.

\section{Motivation}

Firstly, it is important to know that a significant portion of our sample had educational progress delay between 2012 and 2017, as shown in Table \ref{tab:behind_prop}:

\begin{table}[H] \centering 
  \caption{Proportion of students who fell behind in school between 2012 and 2017} 
  \label{tab:behind_prop} 
\begin{tabular}{@{\extracolsep{5pt}} ccc} 
\\[-1.8ex]\hline 
\hline \\[-1.8ex] 
 & Behind & Not Behind \\ 
\hline \\[-1.8ex] 
Proportion & 0.13 & 0.87 \\ 
\hline \\[-1.8ex] 
\end{tabular} 
\end{table} 

Secondly, given the importance that social-emotional and cognitive skills can play in shaping people's paths, especially regarding their success during and after school, we will be inclined to better understand how these skills may relate to falling behind in school, which may be the result of failure or not. In our dataset we have twelve skills that can be considered good or bad by common sense. Two of them can be considered cognitive skills - Literacy and Numeracy\footnote{Measured by language and math scores in a standardized test.} -  and the other ten are social-emotional skills \footnote{We will give a better explanation in Section \ref{sec:data}.}: Assertiveness, Activity, Altruism, Compliance, Order, Self-Discipline, Anxiety, Depression, Aesthetics and Ideas. Two questions that will motivate us from this moment are: (i) How do these skills relate to falling behind in school? (ii) How can we compare the impacts of cognitive and social-emotional skills? In our case, what we will call academic progress delay, or to fall behind in school, is actually the age-grade distortion acquired between 2012 and 2017, i.e. if a student progressed less than five grades in five years (2012 to 2017) we say he/she fell behind or had an academic progress delay. The skills we are using in this paper were measured in 2012 according to some procedures detailed in
Section \ref{sec:data}. In Table \ref{tab:distort_table}, one can see the averages of various skills measured in standard deviations\footnote{The variables were standardized in the sample to have zero mean and unit standard deviation.}, conditioned on the variable 'Behind', which indicates students which fell behind between 2012 and 2017:

% Table created by stargazer v.5.2.2 by Marek Hlavac, Harvard University. E-mail: hlavac at fas.harvard.edu
% Date and time: seg, nov 04, 2019 - 14:19:57
\begin{table}[H] \centering 
  \caption{Comparing the two groups regarding their personal skills} 
  \label{tab:distort_table} 
\begin{tabular}{@{\extracolsep{5pt}} cccc} 
\\[-1.8ex]\hline 
\hline \\[-1.8ex] 
 & Behind & Not Behind & Diff \\ 
\hline \\[-1.8ex] 
Language 2012 & -0.52 & 0.08 & -0.6 \\ 
Mathematics 2012 & -0.44 & 0.07 & -0.51 \\ 
Activity & -0.16 & 0.02 & -0.18 \\ 
Aesthetics & 0.11 & -0.02 & 0.12 \\ 
Altruism & 0.19 & -0.03 & 0.22 \\ 
Anxiety & -0.11 & 0.02 & -0.13 \\ 
Assertiveness & 0.27 & -0.04 & 0.31 \\ 
Compliance & -0.03 & 0 & -0.03 \\ 
Depression & 0.33 & -0.05 & 0.38 \\ 
Ideas & -0.03 & 0 & -0.04 \\ 
Order & 0.16 & -0.02 & 0.19 \\ 
Self-Discipline & 0.05 & -0.01 & 0.05 \\ 
\hline \\[-1.8ex] 
\end{tabular} 
\end{table} 

It can be seen in Table \ref{tab:distort_table} that students who fell behind between 2012 and 2017 have lower average grades in language and mathematics and differ, on average, from the group that did not fall behind with respect to various social-emotional characteristics such as depression - which leads us to believe that among our students, cognitive and social-emotional skills may also be related to whether or not a student fell behind between 2012 and 2017. Although we present a characteristic of the distribution of social-emotional and cognitive skills conditional on the variable 'Behind', we will be more concerned in this paper to estimate the distribution of the 'Behind' variable conditioned to a vector of characteristics, paying more attention to social-emotional and cognitive skills. Therefore, the descriptive analyzes done so far only serves as motivation for a more robust analysis.

\section{Objective}

The main objective of this paper is to better understand how cognitive and social-emotional skills relate to academic progress delay, which may be the result of failure or temporary dropout. For this purpose we use a dataset of approximately 1800 students who studied in the city of Sertãozinho in the countryside of the state of São Paulo and who were in elementary, middle school in 2012 and were re-interviewed in 2017.

\section{Data}\label{sec:data}

The datasets used in this study come from two field surveys conducted by LEPES/USP ("\textit{Laboratório de Estudos e Pesquisas em Economia Social}") in the city of Sertãozinho, in the countryside of the state of São Paulo, in 2012 and 2017. In both years, information was collected about the students' family and socioeconomic context, about the situations they faced in school and about cognitive and social and emotional skills. Since cognitive and social-emotional variables are key variables in our analysis, it is important to explain in detail how they were constructed.

To assess the level of cognitive development of students in 2012 and 2017, we used the result of Language and Mathematics tests prepared by the psychometrist Dr. Ricardo Primi from items available on the platform of the National Institute for Educational Studies and Research Anísio Teixeira (INEP). The test preparation methodology includes the use of the Item Response Theory (IRT), where the final grade is not the number of correct answers, but the student's proficiency level taking into account the difficulty of the items, for example. In 2012, the students, who were in the 5th and 6th grades, answered only one test of each subject. It is important to say that the students who were in the 4th grade in 2012 answered the Big Five Inventory (BFI), discussed bellow, but did not take the exam due some bureaucratic issues.

Regarding the level of social-emotional development, the researchers who designed the research chose to use a well-established social-emotional assessment scale in the literature, which is the Big Five Inventory, or BFI, developed by \cite{john1999big}.
Although, the BFI scale is based on the theory that a person's personality can be roughly described by big five factors, we focused in more recent measures presented in \cite{soto2009ten} which are called 'facets'. We chose to work with facets instead of using the Big Five traits because they are less broad and any result achieved could be interpreted more deeply. The ten available facets and their description, according to \cite{piedmont2013revised}, are:

\begin{enumerate}
    \item \textit{Activity}: High scorers can be described as being energetic, fast-paced, and vigorous. On the other hand, low scorers can be described as unhurried, slow, and deliberate;
    \smallskip
    \item \textit{Aesthetics}: High scorers can be described as those who value aesthetic experiences and who are moved by art and beauty. On the other hand, low scorers can be described as insensitive to art and unappreciative of beauty;
    \smallskip
    \item \textit{Altruism}: High scorers can be described as warm, softhearted, gentle, generous, and kind. On the other hand, low scorers can be described as selfish, cynical, cold, and snobbish;
    \smallskip
    \item \textit{Anxiety}: High scorers can be described as apprehensive, fearful, prone to worry, nervous and tense. On the other hand, low scorers can be described as calm, relaxed, stable, fearless;
    \smallskip
    \item \textit{Assertiveness}: High scorers can be described as dominant, forceful, confident, and decisive. On the other hand, low scorers can be described as unassuming, retiring, and reticent;
    \smallskip
    \item \textit{Compliance}: High scorers can be described as deferential, obliging, and kind. On the other hand, low scorers can be described as stubborn, demanding, headstrong, and hardhearted;
    \smallskip
    \item \textit{Depression}:  High scorers can be described as prone to feelings of guilt, sadness, hopelessness, and loneliness. On the other hand, low scorers can be described as being seldom sad, hopeful, confident, and as feeling worthwhile;
    \smallskip
    \item \textit{Ideas}: High scorers can be described as intellectually curious, analytical, and theoretically oriented. On the other hand, low scorers can be described as pragmatic, factually oriented, and unappreciative of intellectual challenges;
    \smallskip
    \item \textit{Order}: High scorers can be described as precise, efficient, and methodical. On the other hand, low scorers can be described as disorderly, impulsive, and careless;
    \smallskip
    \item \textit{Self-Discipline}: High scorers can be described as organized, thorough, energetic, capable, and efficient. On the other hand, low scorers can be described as unambitious, forgetful, and absent-minded; 
\end{enumerate}

Firstly, we should mention that a higher score in the facets (a.k.a. social-emotional skills) can \textbf{not} always be considered a good thing. Secondly, it is important to say that we do not use the raw scores obtained after applying the scale due to the acquiescence bias. That bias is the tendency of an individual to agree or disagree with statements about his or her attitudes regardless of those \cite{winkler1982controlling}. According to \cite{valentini2017influencia}, if the bias is not corrected the factor estimation may be uncorrelated with the content of the items, which would invalidate the instrument. Given that, all scores obtained for the social-emotional dimensions already discussed were corrected for acquiescence bias.

\subsection{Variables description}

Part of the variables selected to be used in this work was selected from the 2012 dataset, part was selected from the 2017 dataset. More information about our variables:

\begin{itemize}
    \item Grade 2012: Variable that tells us which school grade the student was in 2012;
    \item Semester: Whether the student was born in the first or second semester;
    \item Year: born year of the student (quantitative variable);
    \item White: Binary variable denoting whether the student considers him/herself ethnically as white;
    \item Male: Binary variable denoting whether the student considers him/herself as male (opposed to female);
    \item Pre-K: Binary variable reported by the student whether or not he/she attended pre-kindergarten (only avaliable in 2017);
    \item Kinder: Binary variable reported by the student whether or not he/she attended kindergarten (only avaliable in 2017);
    \item Mother education: variable that tells us the level of formal education of the student's mother. For the sake of inconsistency about what the student and the student's parents were responding to, we will use a way of correcting this that was elaborated by \cite{frank}. Categories: Non-educated (omitted in regression analysis), Elementary, Middle, High, College, Unknown;
    \item School 2012: Categorical variable that tells us whether the student's school in 2012 was state, municipal, federal or private based on the 2012 School Census. Categories: State (omitted in regression analysis), Municipal, Federal or Private;
    \item Failed before 2012: Binary variable reported by the student whether or not he/she failed in school before 2012;
    \item Facets 2012: The social-emotional scores were standardized to have null mean and unit standard deviation in 2012 among those students who form our actual sample;
    \item Language and Mathematics 2012: The tests scores were standardized to have null mean and unit standard deviation in 2012 among those students who form our actual sample;
    \item Behind: Binary variable that tells us if the student fell behind in school between 2012 and 2017 based in his/her grades in both years;

\end{itemize}

In Appendix \ref{sec:descrip} one can see the main descriptive statistics of the variables used in the work.

\subsection{Baseline, attrition and actual sample}

Our baseline of students will consist of the 5th and 6th grade students who were present in 2012 and had no missing values in all variables except the "Pre-K", "Kinder" and "Behind" variables, which depends on the 2017 dataset. In 2012 we have 3223 students who are distributed in 2 school grades: 5th and 6th years. Among these 3223 students, 83,71\% will compose our baseline according to the rule states above. Given that, we have 2698 students in our baseline, as one can check in Table \ref{tab:baseline}: 

% Table generated by Excel2LaTeX from sheet 'Descritivas'
\begin{table}[H]
  \centering
  \begin{threeparttable}
  \caption{Composition of the baseline}
    \begin{tabular}{c|cc|c}
    \toprule
           & 5th grade & 6th grade & Total \\
    \midrule 
    Total  & 1680   & 1543   & 3223 \\
    Baseline  & 1420   & 1278   & 2698 \\ 
    \midrule
    \% Baseline & 84,52\% & 82,82\% & 83,71\% \\
    \bottomrule
    \end{tabular}%
  \label{tab:baseline}%

 \end{threeparttable}
\end{table}%

It is important to state that our baseline is \textbf{not} our actual sample for this study for one major reason: we do not have data for some students in 2017, mainly because we could not reach them in the 2017 field research. Thus, we will use then the amount of 1780 students, which is 65,97\% of the initial amount. Table \ref{tab:actual} provides more details about all this information:

% Table generated by Excel2LaTeX from sheet 'Descritivas'
\begin{table}[H]
  \centering
  \begin{threeparttable}
  \caption{Composition of the actual sample}
    \begin{tabular}{c|cc|c}
    \toprule
           & 5th grade & 6th grade & Total \\
    \midrule 
    Baseline &  1420   & 1278   & 2698 \\
    Actual sample  & 959   & 821   & 1780 \\ 
    \midrule
    \% Actual sample & 67,53\% & 64,24\% & 65,97\% \\
    \bottomrule
    \end{tabular}%
  \label{tab:actual}%

 \end{threeparttable}
\end{table}%

The problem of losing part of the sample between two time points is called attrition. When the attrition is random, that is, when the loss is independent of the characteristics of the students, we do not have many problems. However, when we lose part of the sample in a non-random way, it may be that our results are not extendable to the population of interest, that is all students from the 5th and 6th grades of the Sertãozinho school network in the countryside of the state of São Paulo. If one check in Appendix \ref{sec:attrition}, one will see that the loss was not random: attrition was primarily related to students with less-educated mothers and those with the more problems in school.

Despite a random loss being a sufficient condition for an analysis with extendable results for the population, it is not a necessary condition in our case - I will briefly explain why. Suppose that $Y$ is a binary variable that indicates academic progress delay between 2012 and 2017 for an individual and $X$ a vector of characteristics of the same individual - draws of $X$ are observable in our data. Note that our interest is to better understand the following quantity $\mathbb{P} (Y = 1 | X = x)$, however, we can only estimate $\mathbb{P}(Y = 1 | X = x, A = 0)$ with our data, if $A$ is a variable that indicates attrition. Now imagine that a $Z$ variable vector is the only common cause of falling behind and attrition, and that $U$ and $V$ are other variables that impacts on progress delay or attrition as illustrated in the following causal DAG (Directed Acyclic Graph):

\begin{figure}[H] 
\centering
\includegraphics[width=0.11\textwidth]{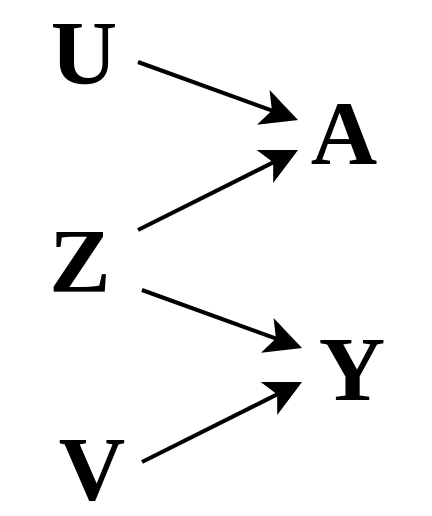} 
\caption[Caption for LOF]{Attrition DAG}
\label{fig:attrition}
\end{figure}

If the DAG is valid and $Z$ is a vector of variables included in $X$, then, given $X=x$, $A$ and $Y$ are conditionally independent and $\mathbb{P}(Y = 1 | X = x) = \mathbb{P}(Y = 1 | X = x, A = 0)$ \cite{pearl2009causal}. It is reasonable to think that this condition is valid since our dataset is rich in demographic, school and personal characteristics of students. On the other hand, if the condition is not valid, we may think that our results will be extendable to a portion of the population, and it is still possible to speculate possible outcomes according to the differences shown in Appendix \ref{sec:attrition}. Although it is not possible to estimate $\mathbb{P}(Y = 1 | X = x)$ directly, we will use this notation from now on taking our sample as given.

\section{Methodology}

\subsection{The model}

The model we will use in our analysis is the Bayesian Logistic Regression model. Using that model we want to directly model the logarithm of students' odds of falling behind in school given their characteristics using a linear predictor. If $Y_i$ is a variable that indicates academic progress delay between 2012 and 2017 for student $i$, $x_i^\top = (1 ~ x_{i1} ~ ... ~ x_{ik})$ is a feature vector for that student, $\theta^\top = (\theta_{0} ~ \theta_{1} ~ ... ~ \theta_{k})$ a parameter vector that helps us to parameterize our model we define:

\begin{align*}
    \pi(x_i,\theta)=\mathbb{P}(Y_i=1|X_i=x_i,\theta)
\end{align*}

Since $\pi (x_i, \theta)$ is the probability of falling behind conditional on the characteristics $ x_i$ and on the $\theta$ parameters, we have that the log of the $ i $ student's odds to have a progress delay between 2012 and 2017, given $\theta$, is:

\begin{equation}
     \text{log}\Big[O(x_i,\theta)\Big]=\text{log}\Bigg [\frac{\pi(x_i,\theta)}{1-\pi(x_i,\theta)}\Bigg ]= x_i^\top \theta
     \label{eq:log_ratio}
\end{equation}

Although logistic regression models the log of the odds, we can obtain $\pi(x_i, \theta)$ and get the conditional probability of progress delay as follows:

\begin{align*}
\pi(x_i,\theta)=\sigma \big( x_i^\top \theta\big)=\frac{1}{1+\text{exp}(-x_i^\top \theta)}
\end{align*}

Where $\sigma (.) $ represents the Sigmoid function (or standard logistic distribution function).

\subsection{Model Parameter Inference}

In the Bayesian framework, it is common to infer about parameters as follows: (i) we adopt an prior distribution for the parameter vector, which captures our knowledge of the quantities of interest before looking at the data; (ii) after observing the data, we update our knowledge about the parameter vector by applying the Bayes Theorem.

\subsubsection{Prior and posterior distributions}

At first we assumed no correlation between the $\theta$ coordinates, and for each of the inputs we assumed a prior Normal distribution $N(0,1000)$, fulfilling the idea of being a weakly informative prior distribution. Our primary objective is to obtain the posterior distribution of the parameter vector, which is given by Equation \ref{eq: post}:

\begin{equation} \label{eq: post}
    p(\theta| Y_{1:n}= y_{1:n},X_{1:n}=x_{1:n})=\frac{\mathbb{P}(Y_{1:n}= y_{1:n}|X_{1:n}=x_{1:n},\theta)p(\theta)}{\int \mathbb{P}(Y_{1:n}= y_{1:n}|X_{1:n}=x_{1:n},\theta)p(\theta) \text{d}\theta}
\end{equation}

Given $Y_{1:n}$ is a vector and $ X_{1:n} $ is a matrix for our entire actual sample ($n$ is the sample size). The major obstacle to calculate the posterior distribution analytically is the calculation of the integral in the denominator, which is intractable. Given that, we resort to a variation of the Hamiltonian Monte Carlo (HMC) \cite{betancourt2017conceptual} algorithm so we can sample from the posterior even without knowing its closed form. That variation of the algorithm is called "No U-Turn Samples" and it is implemented in "RStanarm" package \footnote{\url{https://cran.r-project.org/web/packages/rstanarm/index.html} - accessed in 11/11/2019} in R - that package is built upon "RStan". In our sampling process, we sampled 4 chains, each of which had a burn-in period equals $1000$, a thinning of $100$ and a size of $2500$ (we have $ L =  10000$ samples in total). We conducted a convergence diagnosis of MCMC which can be viewed in more details in Appendix \ref{sec:convergence}. After sampling a reasonable number of times from the posterior distribution, we can move on to the next step. 

\subsection{Analysis of Results}

\subsubsection{Goodness of fit}\label{sec:fit}

In order to understand whether the logistic regression model is a good model for our problem, we will apply a simple cross-validation procedure (one training set and one testing set) for comparing the Receiver Operating Characteristic (ROC) curves and AUC metric (Area Under Curve) between our model and a tuned Random Forest benchmark model \footnote{The model will be tuned in a cross-validation 3-Fold procedure varying the (i) number of trees in the list c(50, 100, 150, 200, 250, 300, 400, 500, 600), (ii) the number of features used in the list c(2,3,4,5,6, 7), (iii) minimum size of nodes in c(1, 3, 5, 10, 15, 20, 30), (iv) sample fraction used in bootstrap in c(.5, .6, .8, 1) and (v) replacement in sampling in c(TRUE, FALSE).}, which has high predictive power. To estimate the conditional probability of falling behind in school using bayesian logistic regression, we will use the concept of posterior predictive distribution, which is calculated as follows:

\begin{align*}
    \mathbb{P}(Y_i=1| X_i= x_i)&=\int \mathbb{P}(Y_i=1| X_i= x_i,\theta)p(\theta | Y_{1:n}= y_{1:n},X_{1:n}=x_{1:n})\textup{d}\theta \\
    &=\mathbb{E}_{\theta|Y_{1:n},X_{1:n}}\Big[ \mathbb{P}(Y_i=1| X_i= x_i,\theta) \Big | Y_{1:n}=y_{1:n},X_{1:n}=x_{1:n}\Big]
\end{align*}

From this moment, we can interpret $ i $ as an out-of-sample individual. As long as we expect to independently sample $\theta$ from its posterior distribution, we may resort to the Law of Large Numbers to approximate the above integral by the following mean:

\begin{align*}
   \mathbb{P}(Y_i=1| X_i= x_i) & \approx \frac{1}{L}\sum_{l=1}^{L}\mathbb{P}(Y_i=1|X_i=x_i,\theta^{(l)}) \\
    &=\frac{1}{L}\sum_{l=1}^{L}\sigma \big( x_i^\top \theta^{(l)}\big)
\end{align*}

Where $\theta \sim p (. | Y_{1:n} = y_{1:n}, X_{1:n} = x_{1:n}) $, i.e., it is sampled from its posterior in a total $ L $ times. Regarding the methodology of comparison between two classifiers, we chose the ROC curves and the AUC metric because they do not depend on the cutoff threshold for classification and are easily interpretable: (i) the ROC curves graphically give us a tradoff between True Positive Rate and True Negative Rate of a binary classifier and (ii) the AUC metric is actually equivalent to the probability that a binary classifier will rank higher an instance of type '1' compared to an instance of type '0', chosen at random \cite{fawcett2006introduction}.

\subsubsection{Odds analysis}\label{sec:odds}

The first analysis we will conduct is direct from obtaining the posterior distribution of the parameter vector. Given the property of the linearity seen in Equation \ref{eq:log_ratio}, we have that the percentage change due to the conditional odds of falling behind due to a $\delta$ change in $x_{ij}$, keeping all other variables constant, is given by:

\begin{align*}
    \Delta_\delta \% O(x_{ij},\theta)=\frac{O(x_{ij}+\delta,\theta)}{O(x_{ij},\theta)}-1=\text{exp}\big(\delta * \theta_{j} \big)-1
\end{align*}

It is important to note that one important implication of linearity of the predictor is that the amount $\Delta_\delta \% O(x_{ij}, \theta) $ does not depend on $ i $, but only on $ \theta_ {j} $ and on $ \delta $. In our analysis, we will consider $\delta = 1 $, which is natural for both binary variables and quantitative variables that are measured in standard deviations. Considering $\delta = 1 $, we have:

\begin{align*}
   \Delta_1 \% O(x_{.j},\theta)=\text{exp}\big(\theta_{j})-1
\end{align*}

Remember that because $ \theta_ {j} $ is a random variable, $\Delta_1 \% O (x_{.j}, \theta) $ is also a random variable and that is why we will examine its distribution directly.

\bigskip

\subsubsection{Testing the importance of each skill variables}\label{sec:hip}

Our analysis regarding the importance of each of the variables in predicting progress delay in school will be analyzed in the light of a hypothesis test that we will propose inspired by \cite{esteves2019pragmatic}. In the proposed framework we have three possible hypotheses for the importance of variable $j$:

\begin{itemize}
    \item $H_0$: the $ j $ variable is not important in determining progress delay in school;
    \item $H_{-}$: the $ j $ variable is negatively important in determining progress delay in school;
    \item $H_{+}$: the $ j $ variable is positively important in determining progress delay in school;
\end{itemize}

Choosing one of the three options will be given by a decision problem under uncertainty. Recalling that $ \theta_ {j}$ assumes values in $ \Theta_j = \mathbb{R} $, we want our hypothesis to be as follows:

\begin{equation}
    \left\{\begin{matrix}
H_0:\theta_{j} \in \left [  \varepsilon_{1},~\varepsilon_{2} \right ]~~~\\ 
\\
H_{-}:\theta_{j} \in \left (-\infty,~\varepsilon_{1} \right )\\ 
\\
H_{+}:\theta_{j} \in \left (  \varepsilon_{2},~\infty \right )~~
\end{matrix}\right.
\end{equation}

Given $\varepsilon_{1}<0$ and $\varepsilon_{2}>0$. The problem with this approach is that it is not straightforward to find $ \varepsilon_ {1} $ and $ \varepsilon_ {2} $ values that make sense. To make this task easier and the result more interpretable, first realize that $ \Delta_1 \% O (x _ {. j}, \theta) = \text {exp} \big (\theta_ {j}) - 1 $ is an increasing function in $ \theta_ {j} $ and with range equals to $ \Omega_j = (- 1, \infty) $. Equivalently, we can rewrite the hypothesis as follows:

\begin{equation}
    \left\{\begin{matrix}
H_0:\Delta_1 \% O(x_{.j},\theta) \in \left [  \varepsilon_{1}',~\varepsilon_{2}' \right ]=\Omega_{j0}~~~\\ 
\\
H_{-}:\Delta_1 \% O(x_{.j},\theta) \in \left (-1,~\varepsilon_{1}' \right )=\Omega_{j-}\\ 
\\
H_{+}:\Delta_1 \% O(x_{.j},\theta) \in \left (  \varepsilon_{2}',~\infty \right )=\Omega_{j+}~~
\end{matrix}\right.
\end{equation}

In the way we lastly presented the hypotheses, one can realize it is easier to choose values of $\varepsilon_ {1}'$ and $ \varepsilon_ {2}' $ that make sense. To define these values, let us remember that the percentage of students who fell behind in school between 2012 and 2017 is $ p = 0.1303 $, so the odds of delay is given by $ \frac{p}{1-p} = 0.1499$. If $ p $ is added by $ d $ points (e.g. $ d = \pm 0.01 $) we get the odds to be $ \frac{p + d}{1- (p + d)} $ and the percentage change in the odds will be given by:

\begin{align}
    \omega_d =  \frac{\frac{p+d}{1-(p+d)}}{\frac{p}{1-p}}-1
\end{align}

Given constants $d_1<0$ and $d_2=-d_1$, we define $ \varepsilon_ {1} '= \omega_{d_1} $ and $ \varepsilon_{2}' = \omega_{d_2} $. The rationale can be elucidated with the following example. Suppose $ d_1 = -0.01 $, so $ \varepsilon_{1} '= - 8.72 \% $ represents the percentage change in the odds of progress delay in school, based on our sample, due to the decreased probability of delay $ p $ by one percentage point. If $ d_2 = 0.01 $, then $ \varepsilon_{2} '= 8.92 \% $ represents the percentage change in odds of progress delay in school, based on our sample, due to the increased probability of delay $ p $ by one percentage point. In a way, the value of $ d_1 $ and hence the value of $ d_2 $ is still somewhat arbitrary, so in our analysis we will assume that $ d_1 $ and $ d_2 $ assume values in $\left \{\pm 0.01, \pm 0.02, \pm 0.03, \pm 0.04, \pm 0.05 \right \} $ to see what our decision would be like in several scenarios.

What remains to be understood is how our decision will be made after we set $ d_1 $ and $ d_2 $. If we define a constant loss function for each of the possible decision errors, i.e. there is no reason to believe that we should value differently different types of errors, our decision to $ \Delta_1 \% O (x _{. j}, \theta) $ with respect to the hypotheses $ H_0, ~ H _- $ and $ H _ + $ is given by:

\begin{equation}
    c_j^*=\underset{c \in \left \{ 0,-,+ \right \}}{\text{argmax}}~\mathbb{P}\Big(\Delta_1 \% O(x_{.j},\theta) \in \Omega_{jc} \Big)
\end{equation}

\subsubsection{Marginal effects analysis}\label{sec:margins}

%We are interested in calculating the average effect of a marginal increase in each of the skills on the predicted probability of school delay for an individual. That is, if $ x_{ij} = \text {skill} _ {ij}, ~ j \in \{1, ..., 12 \} $ denotes the value of $ j $ -th skill for $ i $ -th individual, we want to calculate the amount:

%\begin{align*}
%\mathbb{E}_{\theta}\Bigg[\frac{\partial \mathbb{P}(Y_i=1|X_i=x_i,\theta)}{\partial x_{ij}} \Bigg]
%\end{align*}

%The quantity above is the Bayes Estimator with respect to the quadratic loss for the rate of change in conditional probability varying the $ j $ characteristic of the individual $ i $. By sampling $ L = 10000 $ from the posterior of $ \theta $, we can approximate our quantity of interest as follows:

%\begin{align*}
%\mathbb{E}_{\theta}\Bigg[\frac{\partial \mathbb{P}(Y_i=1|X_i=x_i,\theta)}{\partial x_{ij}} \Bigg] &\approx \frac{1}{L}\sum_{l=1}^{L}\frac{\partial \mathbb{P}(Y_i=1|X_i=x_i, \theta^{(l)})}{\partial x_{ij}}\\
%&=\frac{1}{L}\sum_{l=1}^{L}\frac{\partial \sigma \big( x_i^\top \theta^{(l)}\big)}{\partial x_{ij}}\\
%&=\frac{1}{L}\sum_{l=1}^{L}\sigma \big( x_i^\top \theta^{(l)}\big)\Big[1- \sigma\big( x_i^\top \theta^{(l)}\big)\Big]\theta^{(l)}_{j}
%\end{align*}

%Note that the average marginal effects depends on the characteristics of each individual, so we will analyse the distribution of average marginal effects across the samples.

We are interested in calculating the effect of a marginal increase in each of the skills on the predicted probability of school delay for an individual. That is, if $ x_{ij} = \text {skill} _ {ij}, ~ j \in \{1, ..., 12 \} $ denotes the value of $ j $-th skill for $ i $-th individual (out of sample) we want to calculate the amount:

\begin{align*}
\frac{\partial \mathbb{P}(Y_i=1|X_i=x_i)}{\partial x_{ij}}
\end{align*}

By sampling $ L = 10000 $ from the posterior of $ \theta $, we can approximate our quantity of interest as follows:

\begin{align*}
\frac{\partial \mathbb{P}(Y_i=1|X_i=x_i)}{\partial x_{ij}} &= \frac{\partial }{\partial x_{ij}} \mathbb{E}_{\theta|Y_{1:n},X_{1:n}}\Big[ \mathbb{P}(Y_i=1| X_i= x_i,\theta) \Big | Y_{1:n}=y_{1:n},X_{1:n}=x_{1:n}\Big] \\
&\approx \frac{\partial }{\partial x_{ij}}\frac{1}{L}\sum_{l=1}^{L}\mathbb{P}(Y_i=1|X_i=x_i,\theta^{(l)})\\
&=\frac{1}{L}\sum_{l=1}^{L}\frac{\partial \mathbb{P}(Y_i=1|X_i=x_i, \theta^{(l)})}{\partial x_{ij}}\\
&=\frac{1}{L}\sum_{l=1}^{L}\frac{\partial \sigma \big( x_i^\top \theta^{(l)}\big)}{\partial x_{ij}}\\
&=\frac{1}{L}\sum_{l=1}^{L}\sigma \big( x_i^\top \theta^{(l)}\big)\Big[1- \sigma\big( x_i^\top \theta^{(l)}\big)\Big]\theta^{(l)}_{j}
\end{align*}

Note that the marginal effects depends on the characteristics of each individual, so we will analyse the distribution of the marginal effects across our sample.

\section{Results}

\subsection{Posterior distribution analysis}

The first step in understanding our results is to analyze the posterior distribution of parameters related to skill variables \footnote{Complete results can be found in Appendix \ref{sec:apend_post}}. In order to make the results more interpretable and easy to understand, let us look at the marginal posterior distributions, not taking in consideration the possible relationships between the parameters:

% Table created by stargazer v.5.2.2 by Marek Hlavac, Harvard University. E-mail: hlavac at fas.harvard.edu
% Date and time: sáb, nov 16, 2019 - 01:54:53
\begin{table}[H] \centering 
  \caption{HPD interval for marginal posterior distributions} 
  \label{tab:post_hpd} 
\begin{tabular}{@{\extracolsep{5pt}} ccc} 
\\[-1.8ex]\hline 
\hline \\[-1.8ex] 
 & Lower Bound - HPD 95\% & Upper Bound - HPD 95\% \\ 
\hline \\[-1.8ex] 
Language 2012 & -0.44 & -0.08 \\ 
Mathematics 2012 & -0.58 & -0.17 \\ 
Activity & -0.22 & 0.07 \\ 
Aesthetics & -0.11 & 0.19 \\ 
Altruism & -0.21 & 0.1 \\ 
Anxiety & -0.19 & 0.09 \\ 
Assertiveness & 0.1 & 0.41 \\ 
Compliance & -0.19 & 0.11 \\ 
Depression & 0.16 & 0.45 \\ 
Ideas & -0.06 & 0.23 \\ 
Order & -0.06 & 0.23 \\ 
Self-Discipline & -0.19 & 0.11 \\ 
\hline \\[-1.8ex] 
\end{tabular} 
\end{table} 

In Table \ref{tab:post_hpd} one can see the HPD (Highest Posterior Density) intervals for the posterior distributions of our parameters - we highlight the fact that all parameters of our interest had a unimodal marginal posterior distribution. An interesting point to note is that only four of the marginal distinctions of the parameters of interest had HPD intervals that did not include zero - this gives us clues as to which may be the most important variables in predicting progress delay in school. Two of the variables we have more evidence that can predict our variable of interest have a negative impact on progress delay in school (language and math scores) and two other variables we have the most evidence that can explain our variable of interest have a positive impact on progress delay in school (depression and assertiveness scores). We will put more effort to understand the importance of variables in the next sections. In Table \ref{tab:post_stats} one can see the main statistics for marginal posterior distributions:

% Table created by stargazer v.5.2.2 by Marek Hlavac, Harvard University. E-mail: hlavac at fas.harvard.edu
% Date and time: dom, nov 17, 2019 - 22:01:39
\begin{table}[H] \centering 
  \caption{Basic statistics for marginal posterior distributions} 
  \label{tab:post_stats} 
\begin{tabular}{@{\extracolsep{5pt}} ccccc} 
\\[-1.8ex]\hline 
\hline \\[-1.8ex] 
 & Mean & 1st Qu. & Median & 3rd Qu. \\ 
\hline \\[-1.8ex] 
Language 2012 & -0.27 & -0.33 & -0.27 & -0.21 \\ 
Mathematics 2012 & -0.38 & -0.45 & -0.37 & -0.31 \\ 
Activity & -0.08 & -0.13 & -0.08 & -0.03 \\ 
Aesthetics & 0.04 & -0.01 & 0.04 & 0.09 \\ 
Altruism & -0.05 & -0.11 & -0.05 & 0 \\ 
Anxiety & -0.05 & -0.09 & -0.05 & 0 \\ 
Assertiveness & 0.25 & 0.19 & 0.25 & 0.3 \\ 
Compliance & -0.04 & -0.09 & -0.04 & 0.01 \\ 
Depression & 0.31 & 0.26 & 0.31 & 0.36 \\ 
Ideas & 0.08 & 0.03 & 0.08 & 0.13 \\ 
Order & 0.09 & 0.04 & 0.09 & 0.14 \\ 
Self-Discipline & -0.04 & -0.09 & -0.04 & 0.02 \\ 
\hline \\[-1.8ex] 
\end{tabular} 
\end{table} 

\subsection{Goodness of fit}

A fundamental thing to make our results interesting is the ability of our model to fit well the data we are using. To see how well our model fits the data, we proposed a cross-validation scheme comparing the predictive power of our model with a benchmark model, which in this case will be a Random Forest model. With respect to the Bayesian logistic regression model, we use the predictive distribution of $ Y $ given $ X = x $. In relation to the Random Forest model, we used the combination 'number of trees'=200, 'number of variables per tree'=2, 'minimum node size'=1, 'replacement in sampling'=True, 'fraction sampled in bootstrap'=0.6 chose by means of a grid search in the training set (3-fold CV) in order to maximize the AUC metric. In Figure \ref{fig:roc} one can see a comparison between the two models:

\begin{figure}[H] 
\centering 
\includegraphics[width=0.65\textwidth]{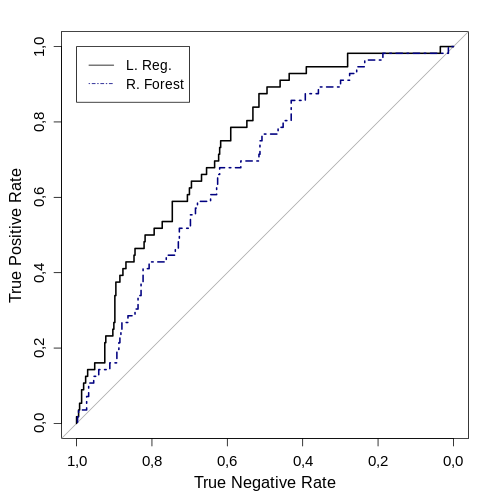} 
\caption[Caption for LOF]{ROC curves - Logistic Regression Vs. Random Forest}
\label{fig:roc}
\end{figure}

Looking at the figure above, it is possible to see that our Bayesian logistic regression model performed better in a classification task when compared to a Random Forest tuned model, which we traditionally consider with great predictive power. We can reiterate our result by looking at Table \ref{tab:auc}:

% Table created by stargazer v.5.2.2 by Marek Hlavac, Harvard University. E-mail: hlavac at fas.harvard.edu
% Date and time: sáb, nov 16, 2019 - 01:31:30
\begin{table}[H] \centering 
    \caption{AUC - Logistic Regression Vs. Random Forest} 
  \label{tab:auc} 
\begin{tabular}{@{\extracolsep{5pt}} cc} 
\\[-1.8ex]\hline 
\hline \\[-1.8ex] 
 & AUC \\ 
\hline \\[-1.8ex] 
Logistic Regression & 0.74 \\ 
Random Forest & 0.68 \\ 
\hline \\[-1.8ex] 
\end{tabular} 
\end{table} 

The great lesson we get from the results seen in this section is that our model could fit well to our data even when compared to more flexible models like a tree ensemble model, that is the random forest.

\subsection{Odds Analysis}

Although we get an important message by looking at the results displayed in Tables \ref{tab:post_hpd} and \ref{tab:post_stats}, interpretation may not be as straightforward as desired, especially when parameters assume larger magnitudes. Thus, using the methodology outlined in Section \ref{sec:odds}, we obtain the distribution of percentage changes in odds (conditional on the $ x $ feature vector) of falling behind in school given an increase in the magnitude of a standard deviation in a skill of interest. In Table \ref{tab:odds_hpd} we can see the HPD intervals of the percentage change distributions in the odds of falling behind in school due to an increase in the magnitude of a standard deviation in a skill of interest:

\begin{table}[H] \centering 
  \caption{HPD interval for percentage change in the odds of study delay (given $x$) from a change in one standard deviation of a skill score}  
  \label{tab:odds_hpd} 
\begin{tabular}{@{\extracolsep{5pt}} ccc} 
\\[-1.8ex]\hline 
\hline \\[-1.8ex] 
 & Lower Bound - HPD 95\% & Upper Bound - HPD 95\% \\ 
\hline \\[-1.8ex] 
Language 2012 & -0.37 & -0.1 \\ 
Mathematics 2012 & -0.44 & -0.16 \\ 
Activity & -0.21 & 0.06 \\ 
Aesthetics & -0.11 & 0.2 \\ 
Altruism & -0.19 & 0.11 \\ 
Anxiety & -0.18 & 0.09 \\ 
Assertiveness & 0.1 & 0.5 \\ 
Compliance & -0.18 & 0.11 \\ 
Depression & 0.17 & 0.56 \\ 
Ideas & -0.07 & 0.25 \\ 
Order & -0.07 & 0.25 \\ 
Self-Discipline & -0.17 & 0.12 \\ 
\hline \\[-1.8ex] 
\end{tabular} 
\end{table} 

An important thing to note is that if the HPD intervals present in Table \ref{tab:post_hpd} do not contain zero, the HPD intervals in Table \ref{tab:odds_hpd} should not contain it either, for a simple mathematical property. Thus, the four skills that bring us the most evidence of their importance are the same as before: language score, math score, depression score and assertiveness score. In this case, for example, an increase in the math score with magnitude of one standard deviation changes the predicted conditional odds of falling behind in school at a magnitude between $ -44 \% $ and $ - 16 \% $ with a probability of $ 0.95 $. Regarding the depression score, an increase in the magnitude of one standard deviation of this score changes the predicted conditional odds of falling behind in school at a magnitude between $ 17 \% $ and $ 56 \% $ with probability of $ 0.95$. The same exercise can be done for other skills. In Table \ref{tab:odds_stats}, we can see some important statistics:

% Table created by stargazer v.5.2.2 by Marek Hlavac, Harvard University. E-mail: hlavac at fas.harvard.edu
% Date and time: sáb, nov 16, 2019 - 02:18:59
\begin{table}[H] \centering 
  \caption{Main statistics for percentage change in the odds of study delay (given $x$) from a change in one standard deviation of a skill score} 
  \label{tab:odds_stats} 
\begin{tabular}{@{\extracolsep{5pt}} ccccc} 
\\[-1.8ex]\hline 
\hline \\[-1.8ex] 
 & Mean & 1st Qu. & Median & 3rd Qu. \\ 
\hline \\[-1.8ex] 
Language 2012 & -0.24 & -0.28 & -0.24 & -0.19 \\ 
Mathematics 2012 & -0.31 & -0.36 & -0.31 & -0.26 \\ 
Activity & -0.07 & -0.12 & -0.08 & -0.03 \\ 
Aesthetics & 0.04 & -0.01 & 0.04 & 0.09 \\ 
Altruism & -0.05 & -0.1 & -0.05 & 0 \\ 
Anxiety & -0.04 & -0.09 & -0.05 & 0 \\ 
Assertiveness & 0.29 & 0.21 & 0.28 & 0.36 \\ 
Compliance & -0.03 & -0.09 & -0.04 & 0.01 \\ 
Depression & 0.36 & 0.29 & 0.36 & 0.43 \\ 
Ideas & 0.09 & 0.03 & 0.08 & 0.14 \\ 
Order & 0.09 & 0.04 & 0.09 & 0.15 \\ 
Self-Discipline & -0.03 & -0.08 & -0.04 & 0.02 \\ 
\hline \\[-1.8ex] 
\end{tabular} 
\end{table} 

The first thing to note is that the odds percentage change distributions are more or less 'balanced' in the sense that the averages approach the medians and the quartiles are more or less symmetrical with respect to the median. One big message we can get from Table \ref{tab:odds_stats} is that, according to the selected statistics, some social-emotional variables can be as important as cognitive skills (math and language) in defining progress delay in school, which is very interesting, given that we traditionally give more importance to the cognitive ones.

\subsection{Testing variables effects}

In the author's point of view, in addition to having an overview of how factors can determine educational progress delay, it is interesting to determine a rule of choice in order to test which factors really matter. In order to make the choices, let's follow the hypothesis testing framework proposed in Section \ref{sec:hip}. As already mentioned, we performed tests for different values of $ d $, which are exposed on the upper horizontal axis of Table \ref{tab:hip}:

% Table created by stargazer v.5.2.2 by Marek Hlavac, Harvard University. E-mail: hlavac at fas.harvard.edu
% Date and time: sáb, nov 16, 2019 - 02:13:28
\begin{table}[H] \centering 
  \caption{Testing variables effects in different scenarios} 
  \label{tab:hip} 
\begin{tabular}{@{\extracolsep{5pt}} cccccc} 
\\[-1.8ex]\hline 
\hline \\[-1.8ex] 
 |d|& 0,01 & 0,02 & 0,03 & 0,04 & 0,05 \\ 
\hline \\[-1.8ex] 
Language 2012 & - & - & 0 & 0 & 0 \\ 
Mathematics 2012 & - & - & - & 0 & 0 \\ 
Activity & 0 & 0 & 0 & 0 & 0 \\ 
Aesthetics & 0 & 0 & 0 & 0 & 0 \\ 
Altruism & 0 & 0 & 0 & 0 & 0 \\ 
Anxiety & 0 & 0 & 0 & 0 & 0 \\ 
Assertiveness & + & + & + & 0 & 0 \\ 
Compliance & 0 & 0 & 0 & 0 & 0 \\ 
Depression & + & + & + & 0 & 0 \\ 
Ideas & 0 & 0 & 0 & 0 & 0 \\ 
Order & + & 0 & 0 & 0 & 0 \\ 
Self-Discipline & 0 & 0 & 0 & 0 & 0 \\  
\hline \\[-1.8ex] 
\end{tabular} 
\end{table} 

In Table \ref{tab:hip} we can see how our decision - regarding the importance of each skill in determining educational progress delay - varies as one chooses different values of $ d $. In the table above, the symbol "0" means that, by adopting variations in percentage points for the prior probability of educational progress delay with magnitude $ | d | $, the variable of interest has no importance in predicting educational progress delay. The symbols "-" and "+" states that the variable of interest is important (negatively and positively) in predicting educational progress delay when adopting variations in percentage points for the prior probability of educational progress delay with magnitude $ | d | $. It is interesting to note that in addition to the four variables we have already mentioned, the social-emotional score of "Order" seems to be related to school delay, which is not very intuitive. However, this importance seems to be less robust when compared to the importance of the other four already mentioned variables.

\subsection{Marginal effects}

Finally, we will present the results regarding the methodology presented in Section \ref{sec:margins}. The marginal effects we calculate give us the rate of change in conditional probability of school delay for an infinitesimal variation in the $ j $ skill score, which is a quantitative variable. As we have seen, this rate of change depends on the characteristics of the individuals in our sample - in the end of the day, we will have a distribution of marginal effects in our sample. In Table \ref{tab:margins_hpd} one can check the HPD intervals for the marginal effects of each variable:

%Although we are presenting the average marginal effects for possible values of $ \theta $, for the sake of clarity and simplicity we will refer to these results only as "marginal effects". 

% Table created by stargazer v.5.2.2 by Marek Hlavac, Harvard University. E-mail: hlavac at fas.harvard.edu
% Date and time: sáb, nov 16, 2019 - 02:23:17
\begin{table}[H] \centering 
  \caption{HPD intervals for marginal effects across samples} 
  \label{tab:margins_hpd} 
\begin{tabular}{@{\extracolsep{5pt}} ccc} 
\\[-1.8ex]\hline 
\hline \\[-1.8ex] 
 & Lower Bound - HPD 95\% & Upper Bound - HPD 95\% \\ 
\hline \\[-1.8ex] 
Portuguese 2012 & -0.06 & 0 \\ 
Mathematics 2012 & -0.08 & 0 \\ 
Activity & -0.02 & 0 \\ 
Aesthetics & 0 & 0.01 \\ 
Altruism & -0.01 & 0 \\ 
Anxiety & -0.01 & 0 \\ 
Assertiveness & 0 & 0.06 \\ 
Compliance & -0.01 & 0 \\ 
Depression & 0 & 0.07 \\ 
Ideas & 0 & 0.02 \\ 
Order & 0 & 0.02 \\ 
Self-Discipline & -0.01 & 0 \\ 
\hline \\[-1.8ex] 
\end{tabular} 
\end{table}

It can be seen in Table \ref{tab:margins_hpd} that the
language and math scores variables and the depression and assertiveness scores continue to stand out from the other variables. Regarding the language score, for example, the result tells us that a change in $ \epsilon $ standard deviations in the score makes us expect a change that can range from $ 0 $ to $ -0.06 * \epsilon $ percentage points in conditional probability of falling behind in school, with a probability of $ 0.95 $. For the result to make sense, $\epsilon $ must be a small number, but even for $ \epsilon = 1 $ we get a reasonable approximation. Regarding the assertiveness score, a variation in $ \epsilon $ standard deviations in the score makes us expect a change that can range from $ 0 $ to $ 0.06 * \epsilon $ percentage points in the conditional probability of falling behind in school, with $ 0.95 $ of probability. In Table \ref{tab:margins_stats}, we have some important statistics:

% Table created by stargazer v.5.2.2 by Marek Hlavac, Harvard University. E-mail: hlavac at fas.harvard.edu
% Date and time: sáb, nov 16, 2019 - 02:23:17
\begin{table}[H] \centering 
  \caption{Statistics for marginal effects across samples} 
  \label{tab:margins_stats} 
\begin{tabular}{@{\extracolsep{5pt}} ccccc} 
\\[-1.8ex]\hline 
\hline \\[-1.8ex] 
 & Mean & 1st Qu. & Median & 3rd Qu. \\ 
\hline \\[-1.8ex] 
Language 2012 & -0.03 & -0.04 & -0.02 & -0.01 \\ 
Mathematics 2012 & -0.04 & -0.05 & -0.03 & -0.02 \\ 
Activity & -0.01 & -0.01 & -0.01 & 0 \\ 
Aesthetics & 0 & 0 & 0 & 0.01 \\ 
Altruism & -0.01 & -0.01 & 0 & 0 \\ 
Anxiety & 0 & -0.01 & 0 & 0 \\ 
Assertiveness & 0.02 & 0.01 & 0.02 & 0.03 \\ 
Compliance & 0 & -0.01 & 0 & 0 \\ 
Depression & 0.03 & 0.01 & 0.03 & 0.04 \\ 
Ideas & 0.01 & 0 & 0.01 & 0.01 \\ 
Order & 0.01 & 0 & 0.01 & 0.01 \\ 
Self-Discipline & 0 & -0.01 & 0 & 0 \\ 
\hline \\[-1.8ex] 
\end{tabular} 
\end{table} 

The results presented in Table \ref{tab:margins_stats} summarize everything presented so far. We emphasize again that an interesting result obtained is that, by comparing the magnitude of the effects, social-emotional skills are as important as cognitive skills in determining educational progress delay.

\section{Conclusion and Discussion}

The main objective of this study was to analyse the relationship between student skills and the fact of falling behind in school, since traditionally in the literature, only the relationship between the contextual variables (family and school) with school delay is evaluated - It is also important to mention the fact that It is hard to have access of a dataset containing a social-emotional assessment of students. To tackle our objective, we used a longitudinal dataset built in two field researches (2012 and 2017) in the city of Sertãozinho, in the countryside of the state of São Paulo. According to the results achieved, we can highlight the following findings, which summarize the contributions of this paper: (i) not all skills are important in determining academic progress delay, in fact, we had only four (out of a dozen) that seem to be more important, namely language and math skills (negative impact on delay) and depression and assertiveness (positive impact on delay); (ii) in terms of magnitude of impact, we can say that social-emotional skills are as important as cognitive skills in determining academic progress delay. 

An additional point that was not covered during the development of this paper is that the results found potentially have a causal interpretation, although I believe that a more detailed study has to be done later. I say this because the context variables are exogenous, we got a great diversity of control variables and, in my view, the only obstacle would be the fact that social-emotional skills have an impact on the probability of falling behind in school through academic performance and did not model that relationship directly. In fact, the hypothesis that social-emotional skills cause cognitive skills is a relevant and supported hypothesis in the literature \cite{cunha2008formulating, cunha2010estimating} - the opposite way is still more questionable. During the experimental phase of this work, I estimated a model without math and language grades, so we would no longer have undesirable control blocking the a path between social-emotional skills and school progress delay. The results were almost identical (posterior distributions) for the social-emotional variables. Because of this, I once again highlight the potential for causal interpretation brought by the results, even if this subject should be the main topic of another paper.

To conclude, I think this work may have an influence on the debate of educational public policies in Brazil and in the world since it deals with current and important issues besides being able to change the way we look at education and what kind of skills we would like to develop in youth. 

\section{Acknowledgments}

I would like to thank CAPES (Higher Education Personnel Improvement Coordination), which is financing my master's degree in statistics at IME/USP and the LEPES/USP (Laboratório de Estudos e Pesquisas em Economia Social) for the dataset used.

\newpage
\bibliographystyle{unsrt}  
\bibliography{template.bib}  %%% Remove comment to use the external .bib file (using bibtex).

@Manual{abntex22013,
  Title                    = {Modelo Canônico de Trabalho Acadêmico com abnTeX2},
  Author                   = {abnTeX2},
  Organization             = {Equipe abnTeX2},
  Year                     = {2013},
  Url                      = {http://abntex2.googlecode.com/}
}

@article{esteves2019pragmatic,
  title={Pragmatic hypotheses in the evolution of Science},
  author={Esteves, Luis Gustavo and Izbicki, Rafael and Stern, Julio Michael and Stern, Rafael Bassi},
  journal={Entropy},
  volume={21},
  number={9},
  pages={883},
  year={2019},
  publisher={Multidisciplinary Digital Publishing Institute}
}

@book{world2015new,
  title={New vision for education: Unlocking the potential of technology},
  author={World Economic Forum},
  year={2015},
  publisher={British Columbia Teachers' Federation}
}

@article{pearl2009causal,
  title={Causal inference in statistics: An overview},
  author={Pearl, Judea and others},
  journal={Statistics surveys},
  volume={3},
  pages={96--146},
  year={2009},
  publisher={The author, under a Creative Commons Attribution License}
}

@article{kankaravs2017personality,
  title={Personality Matters: Relevance and Assessment of Personality Characteristics. OECD Education Working Papers, No. 157.},
  author={Kankara{\v{s}}, Milo{\v{s}}},
  journal={OECD Publishing},
  year={2017},
  publisher={ERIC}
}


@article{robins1998typological,
  title={The typological approach to studying personality},
  author={Robins, Richard W and John, Oliver P and Caspi, Avshalom},
  journal={Methods and models for studying the individual},
  pages={135--160},
  year={1998}
}

@article{bncc,
  title={Secretaria da Educação Básica. Base Nacional Comum
Curricular},
  author={Brasil},
  journal={Ministério da Educação. Brasília – DF},
  year={2018}
}

@article{vehtari2019rank,
  title={Rank-normalization, folding, and localization: An improved $\widehat{R}$ for assessing convergence of MCMC},
  author={Vehtari, Aki and Gelman, Andrew and Simpson, Daniel and Carpenter, Bob and B{\"u}rkner, Paul-Christian},
  journal={arXiv preprint arXiv:1903.08008},
  year={2019}
}

@article{gabry2019visualization,
  title={Visualization in Bayesian workflow},
  author={Gabry, Jonah and Simpson, Daniel and Vehtari, Aki and Betancourt, Michael and Gelman, Andrew},
  journal={Journal of the Royal Statistical Society: Series A (Statistics in Society)},
  volume={182},
  number={2},
  pages={389--402},
  year={2019},
  publisher={Wiley Online Library}
}

@article{betancourt2016diagnosing,
  title={Diagnosing suboptimal cotangent disintegrations in Hamiltonian Monte Carlo},
  author={Betancourt, Michael},
  journal={arXiv preprint arXiv:1604.00695},
  year={2016}
}

@article{frank,
  title={Ambiente familiar e proficiencia escolar: o papel das habilidades
socioemocionais},
  author={Lavinas, João Pedro S. and Costa, Marcos P.C. and Santos, Daniel D.},
  journal={Anais do SJPE&D, Volume 5, Número 5, Santa Maria},
  year={2018}
}

@article{fawcett2006introduction,
  title={An introduction to ROC analysis},
  author={Fawcett, Tom},
  journal={Pattern recognition letters},
  volume={27},
  number={8},
  pages={861--874},
  year={2006},
  publisher={Elsevier}
}

@article{leon2002reprovaccao,
  title={Reprova{\c{c}}{\~a}o, avan{\c{c}}o e evas{\~a}o escolar no Brasil},
  author={Leon, Fernanda Leite Lopez de and Menezes-Filho, Na{\'e}rcio Aquino},
  year={2002},
  publisher={Instituto de Pesquisa Econ{\^o}mica Aplicada (Ipea)}
}

@article{arroyo1995educaccao,
  title={Educa{\c{c}}{\~a}o e exclus{\~a}o da cidadania: In BUFFA},
  author={ARROYO, Miguel G},
  journal={E.: ARROYO, M},
  year={1995}
}

@article{eckstein1999youths,
  title={Why youths drop out of high school: The impact of preferences, opportunities, and abilities},
  author={Eckstein, Zvi and Wolpin, Kenneth I},
  journal={Econometrica},
  volume={67},
  number={6},
  pages={1295--1339},
  year={1999},
  publisher={Wiley Online Library}
}

@article{soares2015fatores,
  title={Fatores associados ao abandono escolar no ensino m{\'e}dio p{\'u}blico de Minas Gerais},
  author={Soares, Tufi Machado and da Silva Fernandes, Neimar and N{\'o}brega, Mariana Calife and Nicolella, Alexandre Chibebe},
  journal={Educa{\c{c}}{\~a}o e Pesquisa},
  volume={41},
  number={3},
  pages={757--772},
  year={2015},
  publisher={Universidade de S{\~a}o Paulo}
}

@book{piedmont2013revised,
  title={The revised NEO Personality Inventory: Clinical and research applications},
  author={Piedmont, Ralph L},
  year={2013},
  publisher={Springer Science \& Business Media}
}

@article{pearl1995causal,
  title={Causal diagrams for empirical research},
  author={Pearl, Judea},
  journal={Biometrika},
  volume={82},
  number={4},
  pages={669--688},
  year={1995},
  publisher={Oxford University Press}
}

@article{denham2012social,
  title={Social--emotional learning profiles of preschoolers' early school success: A person-centered approach},
  author={Denham, Susanne A and Bassett, Hideko and Mincic, Melissa and Kalb, Sara and Way, Erin and Wyatt, Todd and Segal, Yana},
  journal={Learning and individual differences},
  volume={22},
  number={2},
  pages={178--189},
  year={2012},
  publisher={Elsevier}
}

@article{rammstedt2004resilients,
  title={Resilients, overcontrollers, and undercontrollers: The replicability of the three personality prototypes across informants},
  author={Rammstedt, Beatrice and Riemann, Rainer and Angleitner, Alois and Borkenau, Peter},
  journal={European Journal of Personality},
  volume={18},
  number={1},
  pages={1--14},
  year={2004},
  publisher={Wiley Online Library}
}

@article{bohane2017resilients,
  title={Resilients, overcontrollers and undercontrollers: a systematic review of the utility of a personality typology method in understanding adult mental health problems},
  author={Bohane, Laura and Maguire, Nick and Richardson, Thomas},
  journal={Clinical psychology review},
  volume={57},
  pages={75--92},
  year={2017},
  publisher={Elsevier}
}

@article{robins1996resilient,
  title={Resilient, overcontrolled, and undercontrolled boys: Three replicable personality types.},
  author={Robins, Richard W and John, Oliver P and Caspi, Avshalom and Moffitt, Terrie E and Stouthamer-Loeber, Magda},
  journal={Journal of Personality and Social psychology},
  volume={70},
  number={1},
  pages={157},
  year={1996},
  publisher={American Psychological Association}
}

@article{heckman2012hard,
  title={Hard evidence on soft skills},
  author={Heckman, James J and Kautz, Tim},
  journal={Labour economics},
  volume={19},
  number={4},
  pages={451--464},
  year={2012},
  publisher={Elsevier}
}

@mastersthesis{carlos2019papel,
  title={O papel da repet{\^e}ncia escolar sobre vari{\'a}veis de fluxo: uma an{\'a}lise sobre o abandono escolar e chegada ao Ensino M{\'e}dio},
  author={Carlos, Vitor Augusto},
  year={2019},
  school={Universidade de S{\~a}o Paulo}
}


@article{goldenberg2006assessment,
  title={The assessment of emotional intelligence: A comparison of performance-based and self-report methodologies},
  author={Goldenberg, Irina and Matheson, Kimberly and Mantler, Janet},
  journal={Journal of personality assessment},
  volume={86},
  number={1},
  pages={33--45},
  year={2006},
  publisher={Taylor \& Francis}
}

@article{guerra2008linking,
  title={Linking the prevention of problem behaviors and positive youth development: Core competencies for positive youth development and risk prevention},
  author={Guerra, Nancy G and Bradshaw, Catherine P},
  journal={New directions for child and adolescent development},
  volume={2008},
  number={122},
  pages={1--17},
  year={2008},
  publisher={Wiley Online Library}
}

@article{heckman2001importance,
  title={The importance of noncognitive skills: Lessons from the GED testing program},
  author={Heckman, James J and Rubinstein, Yona},
  journal={American Economic Review},
  volume={91},
  number={2},
  pages={145--149},
  year={2001}
}

@article{cunha2008formulating,
  title={Formulating, identifying and estimating the technology of cognitive and noncognitive skill formation},
  author={Cunha, Flavio and Heckman, James J},
  journal={Journal of human resources},
  volume={43},
  number={4},
  pages={738--782},
  year={2008},
  publisher={University of Wisconsin Press}
}
@article{cunha2010estimating,
  title={Estimating the technology of cognitive and noncognitive skill formation},
  author={Cunha, Flavio and Heckman, James J and Schennach, Susanne M},
  journal={Econometrica},
  volume={78},
  number={3},
  pages={883--931},
  year={2010},
  publisher={Wiley Online Library}
}

@misc{miyamoto2015skills,
  title={Skills for social progress: the power of social and emotional skills},
  author={Miyamoto, Koji and Huerta, M and Kubacka, Katarzyna and Ikesako, Hiroko and Oliveira, E},
  year={2015},
  publisher={Paris: OECD, OECD Skills Studies, OECD Center for Research and Innovation (CERI) como parte del proyecto Education and Social Progress (ESP)}
}

@article{durlak2010meta,
  title={A meta-analysis of after-school programs that seek to promote personal and social skills in children and adolescents},
  author={Durlak, Joseph A and Weissberg, Roger P and Pachan, Molly},
  journal={American journal of community psychology},
  volume={45},
  number={3-4},
  pages={294--309},
  year={2010},
  publisher={Springer}
}

@article{covay2010after,
  title={After the bell: Participation in extracurricular activities, classroom behavior, and academic achievement},
  author={Covay, Elizabeth and Carbonaro, William},
  journal={Sociology of Education},
  volume={83},
  number={1},
  pages={20--45},
  year={2010},
  publisher={SAGE Publications Sage CA: Los Angeles, CA}
}

@article{celio2011meta,
  title={A meta-analysis of the impact of service-learning on students},
  author={Celio, Christine I and Durlak, Joseph and Dymnicki, Allison},
  journal={Journal of Experiential Education},
  volume={34},
  number={2},
  pages={164--181},
  year={2011},
  publisher={SAGE Publications Sage CA: Los Angeles, CA}
}

@article{heckman2006effects,
  title={The effects of cognitive and noncognitive abilities on labor market outcomes and social behavior},
  author={Heckman, James J and Stixrud, Jora and Urzua, Sergio},
  journal={Journal of Labor economics},
  volume={24},
  number={3},
  pages={411--482},
  year={2006},
  publisher={The University of Chicago Press}
}
@techreport{kautz2014fostering,
  title={Fostering and measuring skills: Improving cognitive and non-cognitive skills to promote lifetime success},
  author={Kautz, Tim and Heckman, James J and Diris, Ron and Ter Weel, Bas and Borghans, Lex},
  year={2014},
  institution={National Bureau of Economic Research}
}

@article{cunha2007technology,
  title={The technology of skill formation},
  author={Cunha, Flavio and Heckman, James},
  journal={American Economic Review},
  volume={97},
  number={2},
  pages={31--47},
  year={2007}
}

@article{durlak2011impact,
  title={The impact of enhancing students’ social and emotional learning: A meta-analysis of school-based universal interventions},
  author={Durlak, Joseph A and Weissberg, Roger P and Dymnicki, Allison B and Taylor, Rebecca D and Schellinger, Kriston B},
  journal={Child development},
  volume={82},
  number={1},
  pages={405--432},
  year={2011},
  publisher={Wiley Online Library}
}

@Manual{abntex22013b,
  Title                    = {O pacote abntex2cite: Estilos bibliográficos compatíveis com a ABNT NBR 6023},
  Author                   = {abnTeX2 and Lauro César Araujo},
  Organization             = {Equipe abnTeX2},
  Year                     = {2013},
  Url                      = {http://abntex2.googlecode.com/}
}

@Manual{abntex22013c,
  Title                    = {O pacote abntex2cite: tópicos específicos da ABNT NBR 10520:2002 e o estilo bibliográfico alfabético (sistema autor-data)},
  Author                   = {abnTeX2 and Lauro César Araujo},
  Organization             = {Equipe abnTeX2},
  Year                     = {2013},
  Url                      = {http://abntex2.googlecode.com/}
}

@Manual{abntex22013d,
  Title                    = {A classe abntex2: Modelo canônico de trabalhos acadêmicos brasileiros compatível com as normas ABNT NBR 14724:2011, ABNT NBR 6024:2012 e outras},
  Author                   = {abnTeX2 and Lauro César Araujo},
  Organization             = {Equipe abnTeX2},
  Year                     = {2013},
  Url                      = {http://abntex2.googlecode.com/}
}

@book{coelli2005introduction,
  title={An introduction to efficiency and productivity analysis},
  author={Coelli, Timothy J and Rao, Dodla Sai Prasada and O'Donnell, Christopher J and Battese, George Edward},
  year={2005},
  publisher={Springer Science \& Business Media}
}

@article{conroy2008estimation,
  title={An estimation of technical efficiency for Florida public elementary schools},
  author={Conroy, Stephen J and Arguea, Nestor M},
  journal={Economics of Education Review},
  volume={27},
  number={6},
  pages={655--663},
  year={2008},
  publisher={Elsevier}
}

@article{kantabutra2009using,
  title={Using a DEA management tool through a nonparametric approach: an examination of urban-rural effects on Thai school efficiency},
  author={Kantabutra, Sangchan},
  journal={International Journal of Education Policy and Leadership},
  volume={4},
  number={2},
  year={2009}
}


@article{kirjavainen1998efficiency,
  title={Efficiency differences of Finnish senior secondary schools: an application of DEA and Tobit analysis},
  author={Kirjavainen, Tanja and Loikkanent, Heikki A},
  journal={Economics of Education Review},
  volume={17},
  number={4},
  pages={377--394},
  year={1998},
  publisher={Elsevier}
}

@article{delgado2007eficiencia,
  title={Efici{\^e}ncia das escolas p{\'u}blicas estaduais de Minas Gerais},
  author={Delgado, Victor Maia Senna and Machado, Ana Fl{\'a}via},
  year={2007},
  publisher={Instituto de Pesquisa Econ{\^o}mica Aplicada (Ipea)}

@article{fare2000theory,
  title={Theory and application of directional distance functions},
  author={F{\"a}re, Rolf and Grosskopf, Shawna},
  journal={Journal of productivity analysis},
  volume={13},
  number={2},
  pages={93--103},
  year={2000},
  publisher={Springer}
}
}

@article{banker1993maximum,
  title={Maximum likelihood, consistency and data envelopment analysis: a statistical foundation},
  author={Banker, Rajiv D},
  journal={Management science},
  volume={39},
  number={10},
  pages={1265--1273},
  year={1993},
  publisher={INFORMS}
}

@article{banker2008evaluating,
  title={Evaluating contextual variables affecting productivity using data envelopment analysis},
  author={Banker, Rajiv D and Natarajan, Ram},
  journal={Operations research},
  volume={56},
  number={1},
  pages={48--58},
  year={2008},
  publisher={INFORMS}
}

@article{hanushek1979conceptual,
  title={CONCEPTUAL AND EMPIRICAL ISSUES IN THE ESTIMATION OF EDUCATIONAL PRODUCTION FUNCTIONS},
  author={HANUSHEK, ERIC A},
  journal={The Journal of Human Resources},
  volume={14},
  number={3},
  pages={351--388},
  year={1979}
}

@article{hanushek1989impact,
  title={The impact of differential expenditures on school performance},
  author={Hanushek, Eric A},
  journal={Educational researcher},
  volume={18},
  number={4},
  pages={45--62},
  year={1989},
  publisher={Sage Publications}
}

@article{afonso2006cross,
  title={Cross-country efficiency of secondary education provision: A semi-parametric analysis with non-discretionary inputs},
  author={Afonso, Ant{\'o}nio and Aubyn, Miguel St},
  journal={Economic modelling},
  volume={23},
  number={3},
  pages={476--491},
  year={2006},
  publisher={Elsevier}
}

@techreport{cunha2007technology,
  title={The technology of skill formation},
  author={Cunha, Flavio and Heckman, James},
  year={2007},
  institution={National Bureau of Economic Research}
}

@article{heckman2001importance,
  title={The importance of noncognitive skills: Lessons from the GED testing program},
  author={Heckman, James J and Rubinstein, Yona},
  journal={The American Economic Review},
  volume={91},
  number={2},
  pages={145--149},
  year={2001},
  publisher={JSTOR}
}

@article{hanushek1986economics,
  title={The economics of schooling: Production and efficiency in public schools},
  author={Hanushek, Eric A},
  journal={Journal of economic literature},
  volume={24},
  number={3},
  pages={1141--1177},
  year={1986},
  publisher={JSTOR}
}

@article{durlak2011impact,
  title={The impact of enhancing students’ social and emotional learning: A meta-analysis of school-based universal interventions},
  author={Durlak, Joseph A and Weissberg, Roger P and Dymnicki, Allison B and Taylor, Rebecca D and Schellinger, Kriston B},
  journal={Child development},
  volume={82},
  number={1},
  pages={405--432},
  year={2011},
  publisher={Wiley Online Library}
}


@article{conroy2008estimation,
title={An estimation of technical efficiency for Florida public
elementary schools},
  author={Conroy, Stephen J and Arguea, Nestor M},
  journal={Economics of Education Review},
  volume={27},
  number={6},
  pages={655--663},
  year={2008},
  publisher={Elsevier}
}

@techreport{almlund2011personality,
  title={Personality psychology and economics},
  author={Almlund, Mathilde and Duckworth, Angela Lee and Heckman, James J and Kautz, Tim D},
  year={2011},
  institution={National Bureau of Economic Research}
}

@article{lleras2008skills,
  title={Do skills and behaviors in high school matter? The contribution of noncognitive factors in explaining differences in educational attainment and earnings},
  author={Lleras, Christy},
  journal={Social Science Research},
  volume={37},
  number={3},
  pages={888--902},
  year={2008},
  publisher={Elsevier}
}

@article{santos2014desenvolvimento,
  title={Desenvolvimento socioemocional e aprendizado escolar: uma proposta de mensura{\c{c}}{\~a}o para apoiar pol{\'\i}ticas p{\'u}blicas},
  author={SANTOS, Daniel and PRIMI, Ricardo},
  journal={Relat{\'o}rio sobre resultados preliminares do projeto de medi{\c{c}}{\~a}o de compet{\^e}ncias socioemocionais no Rio de Janeiro. S{\~a}o Paulo: OCDE, SEEDUC, Instituto Ayrton Senna},
  year={2014}
}

@article{winkler1982controlling,
  title={Controlling for acquiescence response set in scale development.},
  author={Winkler, John D and Kanouse, David E and Ware, John E},
  journal={Journal of Applied Psychology},
  volume={67},
  number={5},
  pages={555},
  year={1982},
  publisher={American Psychological Association}
}

@misc{trevor2009elements,
  title={The elements of statistical learning: data mining, inference, and prediction},
  author={Trevor, Hastie and Robert, Tibshirani and JH, Friedman},
  year={2009},
  publisher={New York, NY: Springer}
}


@article{tan2013data,
  title={Data mining cluster analysis: basic concepts and algorithms},
  author={Tan, Pang-Ning and Steinbach, Michael and Kumar, Vipin},
  journal={Introduction to data mining},
  year={2013}
}

@article{kodinariya2013review,
  title={Review on determining number of Cluster in K-Means Clustering},
  author={Kodinariya, Trupti M and Makwana, Prashant R},
  journal={International Journal},
  volume={1},
  number={6},
  pages={90--95},
  year={2013}
}

@article{bholowalia2014ebk,
  title={EBK-means: A clustering technique based on elbow method and k-means in WSN},
  author={Bholowalia, Purnima and Kumar, Arvind},
  journal={International Journal of Computer Applications},
  volume={105},
  number={9},
  year={2014},
  publisher={Citeseer}
}

@article{valentini2017influencia,
  title={Influ{\^e}ncia e controle da aquiesc{\^e}ncia na an{\'a}lise fatorial},
  author={Valentini, Felipe},
  journal={Avalia{\c{c}}{\~a}o Psicol{\'o}gica},
  volume={16},
  number={2},
  year={2017},
  publisher={Instituto Brasileiro de Avalia{\c{c}}{\~a}o Psicol{\'o}gica}
}

@article{john1999big,
  title={The Big Five trait taxonomy: History, measurement, and theoretical perspectives},
  author={John, Oliver P and Srivastava, Sanjay},
  journal={Handbook of personality: Theory and research},
  volume={2},
  number={1999},
  pages={102--138},
  year={1999},
  publisher={Guilford}
}

@article{carneiro2007impact,
  title={The impact of early cognitive and non-cognitive skills on later outcomes},
  author={Carneiro, Pedro and Crawford, Claire and Goodman, Alissa},
  year={2007},
  publisher={Centre for Economics of Education}
}

@article{duncan2011nature,
  title={The nature and impact of early achievement skills, attention skills, and behavior problems},
  author={Duncan, Greg J and Magnuson, Katherine},
  journal={Whither opportunity},
  pages={47--70},
  year={2011}
}

@article{mcdonald2008relationship,
  title={The relationship between community violence exposure and mental health symptoms in urban adolescents},
  author={McDonald, Catherine C and Richmond, Therese R},
  journal={Journal of psychiatric and mental health nursing},
  volume={15},
  number={10},
  pages={833--849},
  year={2008},
  publisher={Wiley Online Library}
}

@article{hess2013family,
  title={Family relations, stressful events and internalizing symptoms in adolescence: a longitudinal study},
  author={Hess, Adriana Raquel Binsfeld and Teodoro, Maycoln Leoni Martins and Falcke, Denise},
  journal={The Spanish journal of psychology},
  volume={16},
  pages={E57},
  year={2013},
  publisher={Cambridge Univ Press}
}

@article{del1999conceito,
  title={Conceito e diagn{\'o}stico},
  author={Del Porto, Jos{\'e} Alberto},
  journal={Revista Brasileira de Psiquiatria},
  volume={21},
  pages={06--11},
  year={1999},
  publisher={SciELO Brasil}
}

@article{mello2005curso,
  title={Curso de an{\'a}lise de envolt{\'o}ria de dados},
  author={Mello, JCCBS and Meza, L Angulo and Gomes, Eliane Gon{\c{c}}alves and Neto, L Biondi},
  journal={Simp{\'o}sio Brasileiro de Pesquisa Operacional},
  volume={37},
  pages={2521--2547},
  year={2005}
}

@article{charnes1978measuring,
  title={Measuring the efficiency of decision making units},
  author={Charnes, Abraham and Cooper, William W and Rhodes, Edwardo},
  journal={European journal of operational research},
  volume={2},
  number={6},
  pages={429--444},
  year={1978},
  publisher={Elsevier}
}

@article{banker1984some,
  title={Some models for estimating technical and scale inefficiencies in data envelopment analysis},
  author={Banker, Rajiv D and Charnes, Abraham and Cooper, William Wager},
  journal={Management science},
  volume={30},
  number={9},
  pages={1078--1092},
  year={1984},
  publisher={INFORMS}
}

@article{social2010efficacy,
  title={Efficacy of schoolwide programs to promote social and character development and reduce problem behavior in elementary school children},
  author={Social and Character Development Research Consortium and others},
  journal={Washington, DC: National Center for Education Research, Institute of Education Sciences, US Department of Education},
  year={2010}
}

@article{sklad2012effectiveness,
  title={Effectiveness of School-Based Universal Social, Emotional, and Behavioral Programs: Do They Enhance Students' Development in the Area of Skill, Behavior, and Adjustment?.},
  author={Sklad, Marcin and Diekstra, Rene and De Ritter, Monique and Ben, Jehonathan and Gravesteijn, Carolien},
  journal={Psychology in the Schools},
  volume={49},
  number={9},
  pages={892--909},
  year={2012},
  publisher={ERIC}
}

@article{conceicao2011,
  title={Mensurando a eficiência da atenção primária à saúde em presença de outliers e fatores exógenos: uma análise de dois estágios},
  author={Sousa, Maria da Concei{\c{c}}{\~a}o Sampaio},
  journal={Concurso de Professor Titular da Universidade Federal da Paraíba, na área de Economia do Setor Público},
  year={2011},
}

@article{o1984locus,
  title={Locus of control, work and retirement},
  author={O'Brien, Gordon E},
  journal={Research with the locus of control construct: Extensions and limitations},
  volume={3},
  pages={1--72},
  year={1984},
  publisher={Academic Press New York, NY}
}

@article{milicic2013aprendizagem,
  title={Aprendizagem socioemocional em estudantes de quinta e sexta s{\'e}rie: apresenta{\c{c}}{\~a}o e avalia{\c{c}}{\~a}o de impacto do programa BASE},
  author={Milicic, Neva and Alcalay, Lidia and Berger, Christian and {\'A}lamos, Pilar},
  journal={Revista Ensaio: Avalia{\c{c}}{\~a}o e Pol{\'\i}ticas P{\'u}blicas em Educa{\c{c}}{\~a}o},
  volume={21},
  number={81},
  pages={645--666},
  year={2013}
}

@article{coelho2014impact,
  title={The Impact of a School-Based Social and Emotional Learning Program on the Self-Concept of Middle School Students//O impacto de um programa escolar de Aprendizagem Socioemocional sobre o autoconceito de alunos de 3{\textordmasculine} ciclo},
  author={Coelho, V{\'\i}tor and Sousa, Vanda and Figueira, Ana-Paula},
  journal={Journal of Psychodidactics},
  volume={19},
  number={2},
  year={2014}
}

@article{duckworth2011meta,
  title={A meta-analysis of the convergent validity of self-control measures},
  author={Duckworth, Angela Lee and Kern, Margaret L},
  journal={Journal of Research in Personality},
  volume={45},
  number={3},
  pages={259--268},
  year={2011},
  publisher={Elsevier}
}

Becker, 1967
@book{becker1967human,
title={Human Capital and the Personal Distribution of Income; an
Analytical Approach, by Gary S. Becker},
author={Becker, Gary Stanley},
year={1967},
publisher={Institute of Public Administration}
}

Heckman et al (2006)
@techreport{heckman2006effects,
title={The effects of cognitive and noncognitive abilities on labor
market outcomes and social behavior},
author={Heckman, James J and Stixrud, Jora and Urzua, Sergio},
year={2006},
institution={National Bureau of Economic Research}
}


@article{heckman2001importance,
title={The importance of noncognitive skills: Lessons from the GED
testing program},
author={Heckman, James J and Rubinstein, Yona},
journal={The American Economic Review},
volume={91},
number={2},
pages={145-- 149},
year={2001},
publisher={JSTOR}
}

@techreport{cunha2010investing,
title={Investing in our young people},
author={Cunha, Flavio and Heckman, James J},
year={2010},
institution={National Bureau of Economic Research}
}

(Heckman et al (2012)
@article{heckman2012hard,
title={Hard evidence on soft skills},
author={Heckman, James J and Kautz, Tim},
journal={Labour economics},
volume={19},
number={4},
pages={451-- 464},
year={2012},
publisher={Elsevier}
}

@techreport{barros2015,
title={Apresentação ao EduLab21 - Núcleo de Ciência pela Educação},
author={Barros, Ricardo P},
year={2015},
institution={LANÇAMENTO EDULAB21, LABORATÓRIO DE CIÊNCIAS PARA A EDUCAÇÃO DO INSTITUTO AYRTON SENNA - http://educacaosec21.org.br/wp-content/uploads/2013/08/7.-Roundtable-2_PB_AP2.pdf}
}


@article{roberts2007power, title={The power of personality: The comparative validity of personality traits, socioeconomic status, and cognitive ability for predicting important life outcomes}, author={Roberts, Brent W and Kuncel, Nathan R and Shiner, Rebecca and Caspi,Avshalom and Goldberg, Lewis R}, journal={Perspectives on Psychological Science},volume={2}, number={4}, pages={313-- 345}, year={2007}, publisher={SAGE Publications} }

@article{betancourt2017conceptual,
  title={A conceptual introduction to Hamiltonian Monte Carlo},
  author={Betancourt, Michael},
  journal={arXiv preprint arXiv:1701.02434},
  year={2017}
}

@article{soto2009ten,
  title={Ten facet scales for the Big Five Inventory: Convergence with NEO PI-R facets, self-peer agreement, and discriminant validity},
  author={Soto, Christopher J and John, Oliver P},
  journal={Journal of Research in Personality},
  volume={43},
  number={1},
  pages={84--90},
  year={2009},
  publisher={Elsevier}
}

\begin{thebibliography}{10}

\bibitem{world2015new}
World~Economic Forum.
\newblock {\em New vision for education: Unlocking the potential of
  technology}.
\newblock British Columbia Teachers' Federation, 2015.

\bibitem{heckman2006effects}
James~J Heckman, Jora Stixrud, and Sergio Urzua.
\newblock The effects of cognitive and noncognitive abilities on labor market
  outcomes and social behavior.
\newblock {\em Journal of Labor economics}, 24(3):411--482, 2006.

\bibitem{kautz2014fostering}
Tim Kautz, James~J Heckman, Ron Diris, Bas Ter~Weel, and Lex Borghans.
\newblock Fostering and measuring skills: Improving cognitive and non-cognitive
  skills to promote lifetime success.
\newblock Technical report, National Bureau of Economic Research, 2014.

\bibitem{miyamoto2015skills}
Koji Miyamoto, M~Huerta, Katarzyna Kubacka, Hiroko Ikesako, and E~Oliveira.
\newblock Skills for social progress: the power of social and emotional skills,
  2015.

\bibitem{leon2002reprovaccao}
Fernanda Leite Lopez~de Leon and Na{\'e}rcio~Aquino Menezes-Filho.
\newblock Reprova{\c{c}}{\~a}o, avan{\c{c}}o e evas{\~a}o escolar no brasil.
\newblock 2002.

\bibitem{arroyo1995educaccao}
Miguel~G ARROYO.
\newblock Educa{\c{c}}{\~a}o e exclus{\~a}o da cidadania: In buffa.
\newblock {\em E.: ARROYO, M}, 1995.

\bibitem{eckstein1999youths}
Zvi Eckstein and Kenneth~I Wolpin.
\newblock Why youths drop out of high school: The impact of preferences,
  opportunities, and abilities.
\newblock {\em Econometrica}, 67(6):1295--1339, 1999.

\bibitem{soares2015fatores}
Tufi~Machado Soares, Neimar da~Silva~Fernandes, Mariana~Calife N{\'o}brega, and
  Alexandre~Chibebe Nicolella.
\newblock Fatores associados ao abandono escolar no ensino m{\'e}dio
  p{\'u}blico de minas gerais.
\newblock {\em Educa{\c{c}}{\~a}o e Pesquisa}, 41(3):757--772, 2015.

\bibitem{carlos2019papel}
Vitor~Augusto Carlos.
\newblock O papel da repet{\^e}ncia escolar sobre vari{\'a}veis de fluxo: uma
  an{\'a}lise sobre o abandono escolar e chegada ao ensino m{\'e}dio.
\newblock Master's thesis, Universidade de S{\~a}o Paulo, 2019.

\bibitem{john1999big}
Oliver~P John and Sanjay Srivastava.
\newblock The big five trait taxonomy: History, measurement, and theoretical
  perspectives.
\newblock {\em Handbook of personality: Theory and research}, 2(1999):102--138,
  1999.

\bibitem{soto2009ten}
Christopher~J Soto and Oliver~P John.
\newblock Ten facet scales for the big five inventory: Convergence with neo
  pi-r facets, self-peer agreement, and discriminant validity.
\newblock {\em Journal of Research in Personality}, 43(1):84--90, 2009.

\bibitem{piedmont2013revised}
Ralph~L Piedmont.
\newblock {\em The revised NEO Personality Inventory: Clinical and research
  applications}.
\newblock Springer Science \& Business Media, 2013.

\bibitem{winkler1982controlling}
John~D Winkler, David~E Kanouse, and John~E Ware.
\newblock Controlling for acquiescence response set in scale development.
\newblock {\em Journal of Applied Psychology}, 67(5):555, 1982.

\bibitem{valentini2017influencia}
Felipe Valentini.
\newblock Influ{\^e}ncia e controle da aquiesc{\^e}ncia na an{\'a}lise
  fatorial.
\newblock {\em Avalia{\c{c}}{\~a}o Psicol{\'o}gica}, 16(2), 2017.

\bibitem{frank}
João Pedro~S. Lavinas, Marcos~P.C. Costa, and Daniel~D. Santos.
\newblock Ambiente familiar e proficiencia escolar: o papel das habilidades
  socioemocionais.
\newblock {\em Anais do SJPE \& D, Volume 5, Número 5, Santa Maria}, 2018.

\bibitem{pearl2009causal}
Judea Pearl et~al.
\newblock Causal inference in statistics: An overview.
\newblock {\em Statistics surveys}, 3:96--146, 2009.

\bibitem{betancourt2017conceptual}
Michael Betancourt.
\newblock A conceptual introduction to hamiltonian monte carlo.
\newblock {\em arXiv preprint arXiv:1701.02434}, 2017.

\bibitem{fawcett2006introduction}
Tom Fawcett.
\newblock An introduction to roc analysis.
\newblock {\em Pattern recognition letters}, 27(8):861--874, 2006.

\bibitem{esteves2019pragmatic}
Luis~Gustavo Esteves, Rafael Izbicki, Julio~Michael Stern, and Rafael~Bassi
  Stern.
\newblock Pragmatic hypotheses in the evolution of science.
\newblock {\em Entropy}, 21(9):883, 2019.

\bibitem{cunha2008formulating}
Flavio Cunha and James~J Heckman.
\newblock Formulating, identifying and estimating the technology of cognitive
  and noncognitive skill formation.
\newblock {\em Journal of human resources}, 43(4):738--782, 2008.

\bibitem{cunha2010estimating}
Flavio Cunha, James~J Heckman, and Susanne~M Schennach.
\newblock Estimating the technology of cognitive and noncognitive skill
  formation.
\newblock {\em Econometrica}, 78(3):883--931, 2010.

\bibitem{betancourt2016diagnosing}
Michael Betancourt.
\newblock Diagnosing suboptimal cotangent disintegrations in hamiltonian monte
  carlo.
\newblock {\em arXiv preprint arXiv:1604.00695}, 2016.

\bibitem{gabry2019visualization}
Jonah Gabry, Daniel Simpson, Aki Vehtari, Michael Betancourt, and Andrew
  Gelman.
\newblock Visualization in bayesian workflow.
\newblock {\em Journal of the Royal Statistical Society: Series A (Statistics
  in Society)}, 182(2):389--402, 2019.

\bibitem{vehtari2019rank}
Aki Vehtari, Andrew Gelman, Daniel Simpson, Bob Carpenter, and Paul-Christian
  B{\"u}rkner.
\newblock Rank-normalization, folding, and localization: An improved
  $\widehat{R}$ for assessing convergence of mcmc.
\newblock {\em arXiv preprint arXiv:1903.08008}, 2019.

\end{thebibliography}

\newpage
\section{Appendix}

\subsection{Attrition}\label{sec:attrition}

% Table created by stargazer v.5.2.2 by Marek Hlavac, Harvard University. E-mail: hlavac at fas.harvard.edu
% Date and time: seg, nov 04, 2019 - 14:19:52
\begin{table}[H] \centering 
  \caption{Comparing the two groups} 
   
\begin{tabular}{@{\extracolsep{5pt}} cccc} 
\\[-1.8ex]\hline 
\hline \\[-1.8ex] 
 & Attrition & Non-Attrition & Diff. \\ 
\hline \\[-1.8ex] 
Unknown & 0.06 & 0.01 & 0.05 \\ 
Non-educated & 0.14 & 0.15 & -0.01 \\ 
Elementary & 0.34 & 0.29 & 0.05 \\ 
Middle & 0.19 & 0.22 & -0.03 \\ 
High & 0.19 & 0.26 & -0.07 \\ 
College & 0.08 & 0.07 & 0.01 \\ 
\hline \\[-1.8ex] 
\end{tabular} 
\end{table} 

% Table created by stargazer v.5.2.2 by Marek Hlavac, Harvard University. E-mail: hlavac at fas.harvard.edu
% Date and time: seg, nov 04, 2019 - 14:19:53
\begin{table}[H] \centering 
  \caption{Comparing the two groups} 
   
\begin{tabular}{@{\extracolsep{5pt}} cccc} 
\\[-1.8ex]\hline 
\hline \\[-1.8ex] 
 & Attrition & Non-Attrition & Diff. \\ 
\hline \\[-1.8ex] 
State & 0.31 & 0.27 & 0.04 \\ 
Municipal & 0.62 & 0.63 & -0.01 \\ 
Private & 0.08 & 0.1 & -0.02 \\ 
\hline \\[-1.8ex] 
\end{tabular} 
\end{table} 

% Table created by stargazer v.5.2.2 by Marek Hlavac, Harvard University. E-mail: hlavac at fas.harvard.edu
% Date and time: seg, nov 04, 2019 - 14:19:54
\begin{table}[H] \centering 
  \caption{Comparing the two groups} 
   
\begin{tabular}{@{\extracolsep{5pt}} cccc} 
\\[-1.8ex]\hline 
\hline \\[-1.8ex] 
 & Attrition & Non-Attrition & Diff. \\ 
\hline \\[-1.8ex] 
White & 0.36 & 0.41 & -0.04 \\ 
Man & 0.55 & 0.48 & 0.07 \\ 
Failed before 2012 & 0.47 & 0.24 & 0.23 \\ 
\hline \\[-1.8ex] 
\end{tabular} 
\end{table} 

In Table \ref{tab:att_skills}, we standardize the scores (mean = 0, std = 1) before selecting only those we could find in 2017:

% Table created by stargazer v.5.2.2 by Marek Hlavac, Harvard University. E-mail: hlavac at fas.harvard.edu
% Date and time: seg, nov 04, 2019 - 14:19:54
\begin{table}[H] \centering 
  \caption{Comparing the two groups} 
  \label{tab:att_skills} 
\begin{tabular}{@{\extracolsep{5pt}} cccc} 
\\[-1.8ex]\hline 
\hline \\[-1.8ex] 
 & Attrition & Non-Attrition & Diff. \\ 
\hline \\[-1.8ex] 
Language 2012 & -0.27 & 0.14 & -0.41 \\ 
Mathematics 2012 & -0.23 & 0.12 & -0.35 \\ 
Activity & -0.08 & 0.04 & -0.12 \\ 
Aesthetics & 0.1 & -0.05 & 0.15 \\ 
Altruism & 0.04 & -0.02 & 0.06 \\ 
Anxiety & 0 & 0 & 0 \\ 
Assertiveness & 0.05 & -0.03 & 0.08 \\ 
Compliance & -0.04 & 0.02 & -0.07 \\ 
Depression & 0.07 & -0.04 & 0.11 \\ 
Ideas & -0.04 & 0.02 & -0.05 \\ 
Order & 0.07 & -0.04 & 0.11 \\ 
Self-Discipline & 0.03 & -0.02 & 0.05 \\ 
\hline \\[-1.8ex] 
\end{tabular} 
\end{table}

\subsection{Descriptive statistics}\label{sec:descrip}

% Table created by stargazer v.5.2.2 by Marek Hlavac, Harvard University. E-mail: hlavac at fas.harvard.edu
% Date and time: seg, nov 04, 2019 - 14:19:56
\begin{table}[H] \centering 
  \caption{Descriptive statistics in actual sample}
   
\begin{tabular}{@{\extracolsep{5pt}} cc} 
\\[-1.8ex]\hline 
\hline \\[-1.8ex] 
 & Proportion \\ 
\hline \\[-1.8ex] 
Unknown & 0.01 \\ 
Non-educated & 0.15 \\ 
Elementary & 0.29 \\ 
Middle & 0.22 \\ 
High & 0.26 \\ 
College & 0.07 \\ 
\hline \\[-1.8ex] 
\end{tabular} 
\end{table} 

% Table created by stargazer v.5.2.2 by Marek Hlavac, Harvard University. E-mail: hlavac at fas.harvard.edu
% Date and time: seg, nov 04, 2019 - 14:19:56
\begin{table}[H] \centering 
  \caption{Descriptive statistics in actual sample} 
   
\begin{tabular}{@{\extracolsep{5pt}} cc} 
\\[-1.8ex]\hline 
\hline \\[-1.8ex] 
 & Proportion \\ 
\hline \\[-1.8ex] 
State & 0.27 \\ 
Municipal & 0.63 \\ 
Private & 0.1 \\ 
\hline \\[-1.8ex] 
\end{tabular} 
\end{table} 

% Table created by stargazer v.5.2.2 by Marek Hlavac, Harvard University. E-mail: hlavac at fas.harvard.edu
% Date and time: seg, nov 04, 2019 - 14:19:56
\begin{table}[H] \centering 
  \caption{Descriptive statistics in actual sample} 
   
\begin{tabular}{@{\extracolsep{5pt}} cc} 
\\[-1.8ex]\hline 
\hline \\[-1.8ex] 
 & Proportion \\ 
\hline \\[-1.8ex] 
White & 0.41 \\ 
Man & 0.48 \\ 
Failed before 2012 & 0.24 \\ 
Pre-K & 0.48 \\ 
Kinder & 0.86 \\ 
\hline \\[-1.8ex] 
\end{tabular} 
\end{table} 

% Table created by stargazer v.5.2.2 by Marek Hlavac, Harvard University. E-mail: hlavac at fas.harvard.edu
% Date and time: seg, nov 04, 2019 - 14:19:57
\begin{table}[H] \centering 
  \caption{Descriptive statistics in actual sample} 
   
\begin{tabular}{@{\extracolsep{5pt}} ccccccc} 
\\[-1.8ex]\hline 
\hline \\[-1.8ex] 
 & Min. & 1st Qu. & Median & Mean & 3rd Qu. & Max. \\ 
\hline \\[-1.8ex] 
Portuguese 2012 & -4.04 & -0.48 & 0.11 & 0 & 0.72 & 1.82 \\ 
Mathematics 2012 & -3.34 & -0.64 & 0.03 & 0 & 0.71 & 2.08 \\ 
Activity & -3.5 & -0.7 & 0.04 & 0 & 0.76 & 2.62 \\ 
Aesthetics & -3.45 & -0.64 & -0.05 & 0 & 0.66 & 3.33 \\ 
Altruism & -3.02 & -0.7 & -0.05 & 0 & 0.6 & 3.97 \\ 
Anxiety & -3.87 & -0.42 & 0.01 & 0 & 0.73 & 3.27 \\ 
Assertiveness & -3.21 & -0.62 & -0.05 & 0 & 0.65 & 3.39 \\ 
Compliance & -2.68 & -0.7 & -0.02 & 0 & 0.71 & 2.8 \\ 
Depression & -2.45 & -0.74 & -0.16 & 0 & 0.57 & 5.5 \\ 
Ideas & -4.09 & -0.62 & 0 & 0 & 0.64 & 3.31 \\ 
Order & -2.51 & -0.75 & -0.15 & 0 & 0.64 & 4.37 \\ 
Self-Discipline & -3.13 & -0.63 & 0 & 0 & 0.65 & 3.14 \\ 
\hline \\[-1.8ex] 
\end{tabular} 
\end{table} 

\subsection{Posterior distribution analysis}\label{sec:apend_post}

% Table created by stargazer v.5.2.2 by Marek Hlavac, Harvard University. E-mail: hlavac at fas.harvard.edu
% Date and time: sáb, nov 16, 2019 - 02:06:56
\begin{table}[H] \centering 
  \caption{HPD interval for marginal posterior distributions} 
   
\begin{tabular}{@{\extracolsep{5pt}} ccc} 
\\[-1.8ex]\hline 
\hline \\[-1.8ex] 
 & Lower Bound - HPD 95\% & Upper Bound - HPD 95\% \\ 
\hline \\[-1.8ex] 
Intercept & -11.02 & -4.99 \\ 
Grade 2012 & 0.62 & 1.4 \\ 
Birth Year & 0 & 0 \\ 
Birth Semester & -0.43 & 0.17 \\ 
School 2012: Municipal & 0.49 & 1.38 \\ 
School 2012: Private & 0.74 & 1.94 \\ 
Ethinicity: White & -0.5 & 0.13 \\ 
Gender: Male & 0.16 & 0.79 \\ 
Failed Before 2012 & 0.13 & 0.79 \\ 
Mother Educ.: Unknown & -4.67 & 0.33 \\ 
Mother Educ.: Elementary & 0.01 & 0.93 \\ 
Mother Educ.: Middle & -0.12 & 0.87 \\ 
Mother Educ.: High & -0.53 & 0.52 \\ 
Mother Educ.: College & -1.75 & 0.03 \\ 
Childhood Educ: Pre-k & -0.24 & 0.36 \\ 
Childhood Educ: Kinder & -0.25 & 0.61 \\ 
Language 2012 & -0.44 & -0.08 \\ 
Mathematics 2012 & -0.58 & -0.17 \\ 
Activity & -0.22 & 0.07 \\ 
Aesthetics & -0.11 & 0.19 \\ 
Altruism & -0.21 & 0.1 \\ 
Anxiety & -0.19 & 0.09 \\ 
Assertiveness & 0.1 & 0.41 \\ 
Compliance & -0.19 & 0.11 \\ 
Depression & 0.16 & 0.45 \\ 
Ideas & -0.06 & 0.23 \\ 
Order & -0.06 & 0.23 \\ 
Self-Discipline & -0.19 & 0.11 \\ 
\hline \\[-1.8ex] 
\end{tabular} 
\end{table} 

% Table created by stargazer v.5.2.2 by Marek Hlavac, Harvard University. E-mail: hlavac at fas.harvard.edu
% Date and time: dom, nov 17, 2019 - 22:03:28
\begin{table}[H] \centering 
  \caption{Basic statistics for marginal posterior distributions} 
   
\begin{tabular}{@{\extracolsep{5pt}} ccccc} 
\\[-1.8ex]\hline 
\hline \\[-1.8ex] 
 & Mean & 1st Qu. & Median & 3rd Qu. \\ 
\hline \\[-1.8ex] 
Intercept & -8.05 & -9.09 & -8.03 & -7.02 \\ 
Grade 2012 & 1.01 & 0.88 & 1.01 & 1.15 \\ 
Birth Year & 0 & 0 & 0 & 0 \\ 
Birth Semester & -0.14 & -0.24 & -0.14 & -0.03 \\ 
School 2012: Municipal & 0.94 & 0.78 & 0.94 & 1.09 \\ 
School 2012: Private & 1.33 & 1.13 & 1.34 & 1.54 \\ 
Ethinicity: White & -0.19 & -0.3 & -0.19 & -0.08 \\ 
Gender: Male & 0.47 & 0.35 & 0.47 & 0.58 \\ 
Failed Before 2012 & 0.46 & 0.34 & 0.46 & 0.57 \\ 
Mother Educ.: Unknown & -1.97 & -2.7 & -1.82 & -1.04 \\ 
Mother Educ.: Elementary & 0.46 & 0.29 & 0.46 & 0.62 \\ 
Mother Educ.: Middle & 0.4 & 0.23 & 0.4 & 0.57 \\ 
Mother Educ.: High & -0.01 & -0.19 & -0.01 & 0.18 \\ 
Mother Educ.: College & -0.86 & -1.16 & -0.84 & -0.54 \\ 
Childhood Educ: Pre-k & 0.07 & -0.04 & 0.06 & 0.17 \\ 
Childhood Educ: Kinder & 0.17 & 0.02 & 0.16 & 0.31 \\ 
Language 2012 & -0.27 & -0.33 & -0.27 & -0.21 \\ 
Mathematics 2012 & -0.38 & -0.45 & -0.37 & -0.31 \\ 
Activity & -0.08 & -0.13 & -0.08 & -0.03 \\ 
Aesthetics & 0.04 & -0.01 & 0.04 & 0.09 \\ 
Altruism & -0.05 & -0.11 & -0.05 & 0 \\ 
Anxiety & -0.05 & -0.09 & -0.05 & 0 \\ 
Assertiveness & 0.25 & 0.19 & 0.25 & 0.3 \\ 
Compliance & -0.04 & -0.09 & -0.04 & 0.01 \\ 
Depression & 0.31 & 0.26 & 0.31 & 0.36 \\ 
Ideas & 0.08 & 0.03 & 0.08 & 0.13 \\ 
Order & 0.09 & 0.04 & 0.09 & 0.14 \\ 
Self-Discipline & -0.04 & -0.09 & -0.04 & 0.02 \\ 
\hline \\[-1.8ex] 
\end{tabular} 
\end{table} 

\newpage
\subsection{Testing variables impacts}

% Table created by stargazer v.5.2.2 by Marek Hlavac, Harvard University. E-mail: hlavac at fas.harvard.edu
% Date and time: sáb, nov 16, 2019 - 02:11:25
\begin{table}[H] \centering 
  \caption{Testing variables impacts} 
   
\begin{tabular}{@{\extracolsep{5pt}} cccccc} 
\\[-1.8ex]\hline 
\hline \\[-1.8ex] 
 & 0,01 & 0,02 & 0,03 & 0,04 & 0,05 \\ 
\hline \\[-1.8ex] 
Intercept & - & - & - & - & - \\ 
Grade 2012 & + & + & + & + & + \\ 
Birth Year & 0 & 0 & 0 & 0 & 0 \\ 
Birth Semester & - & 0 & 0 & 0 & 0 \\ 
School 2012: Municipal & + & + & + & + & + \\ 
School 2012: Private & + & + & + & + & + \\ 
Ethinicity: White & - & - & 0 & 0 & 0 \\ 
Gender: Male & + & + & + & + & + \\ 
Failed Before 2012 & + & + & + & + & + \\ 
Mother Educ.: Unknown & - & - & - & - & - \\ 
Mother Educ.: Elementary & + & + & + & + & + \\ 
Mother Educ.: Middle & + & + & + & + & + \\ 
Mother Educ.: High & - & 0 & 0 & 0 & 0 \\ 
Mother Educ.: College & - & - & - & - & - \\ 
Childhood Educ: Pre-k & + & 0 & 0 & 0 & 0 \\ 
Childhood Educ: Kinder & + & + & 0 & 0 & 0 \\ 
Language 2012 & - & - & 0 & 0 & 0 \\ 
Mathematics 2012 & - & - & - & 0 & 0 \\ 
Activity & 0 & 0 & 0 & 0 & 0 \\ 
Aesthetics & 0 & 0 & 0 & 0 & 0 \\ 
Altruism & 0 & 0 & 0 & 0 & 0 \\ 
Anxiety & 0 & 0 & 0 & 0 & 0 \\ 
Assertiveness & + & + & + & 0 & 0 \\ 
Compliance & 0 & 0 & 0 & 0 & 0 \\ 
Depression & + & + & + & 0 & 0 \\ 
Ideas & 0 & 0 & 0 & 0 & 0 \\ 
Order & + & 0 & 0 & 0 & 0 \\ 
Self-Discipline & 0 & 0 & 0 & 0 & 0 \\  
\hline \\[-1.8ex] 
\end{tabular} 
\end{table} 

\newpage
\subsection{Convergence diagnosis of MCMC}\label{sec:convergence}

\subsubsection{Divergent transitions No U-Turn Sampler}

In addition to the general methods for diagnosing convergence of MCMC algorithms, we have some specific methods for the "No U-Turn Sampler" algorithm. The Stan package automatically offers ways to infer about method convergence. The most straightforward way is to check graphically if the package reports "Divergent transitions"\cite{betancourt2016diagnosing, betancourt2017conceptual, gabry2019visualization}. According to the package documentation itself \footnote{\url{https://mc-stan.org/docs/2_19/reference-manual/divergent-transitions}}: "A divergence arises when the simulated Hamiltonian trajectory departs from the true trajectory as measured by departure of the Hamiltonian value from its initial value. When this divergence is too high, the simulation has gone off the rails and cannot be trusted". In the figure below we have that every tick represents a parameter on the horizontal axis and each line represents an isolated sample. One can see that the package did not identify divergent transitions compared to the examples in \cite{gabry2019visualization}:

\begin{figure}[H] 
\centering 
\includegraphics[width=.9\textwidth]{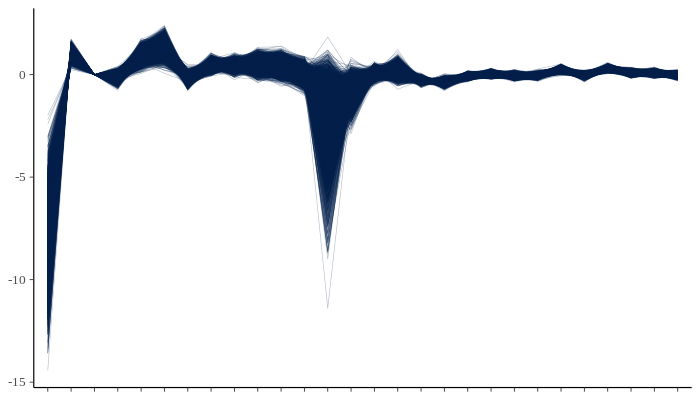} 
\caption[Caption for LOF]{Divergent transitions No U-Turn Sampler}
\end{figure}

\newpage
\subsubsection{$\hat{R}$ and effective sample size}

$ \hat{R} $ is a classic diagnosis statistic for multiple chains of MCMC algorithms. A new version is presented in \cite{vehtari2019rank} and corrects some points made by the authors. If $ \hat{R} $ is too close to 1, we have evidence that the chains have converged to the same distribution. In addition to $ \hat{R} $, in the table below we can check the effective sample size, which is very close to $100\%$ for all variables, indicating that autocorrelations within chains are low. Then, we have two more evidences that we sampled from the distributions of interest:

% Table created by stargazer v.5.2.2 by Marek Hlavac, Harvard University. E-mail: hlavac at fas.harvard.edu
% Date and time: qua, nov 13, 2019 - 16:30:50
\begin{table}[H] \centering 
  \caption{$\hat{R}$ and effective sample size} 
   
\begin{tabular}{@{\extracolsep{5pt}} cccc} 
\\[-1.8ex]\hline 
\hline \\[-1.8ex] 
 & Rhat & Bulk\_ESS & Tail\_ESS \\ 
\hline \\[-1.8ex] 
Intercept & 1 & 4994 & 4928 \\ 
Grade 2012 & 1 & 4852 & 4839 \\ 
Birth Year & 1 & 5111 & 5008 \\ 
Birth Semester & 1 & 4639 & 4872 \\ 
School 2012: Municipal & 1 & 5190 & 5031 \\ 
School 2012: Private & 1 & 4823 & 4648 \\ 
Ethinicity: White & 1 & 5084 & 4911 \\ 
Gender: Male & 1 & 4856 & 4780 \\ 
Failed Before 2012 & 1 & 5142 & 4724 \\ 
Mother Educ.: Unknown & 1 & 5094 & 4391 \\ 
Mother Educ.: Elementary & 1 & 4849 & 4910 \\ 
Mother Educ.: Middle & 1 & 4836 & 4907 \\ 
Mother Educ.: High & 1 & 4796 & 4400 \\ 
Mother Educ.: College & 1 & 5050 & 4870 \\ 
Childhood Educ: Pre-k & 1 & 5016 & 5119 \\ 
Childhood Educ: Kinder & 1 & 5125 & 4758 \\ 
Language 2012 & 1 & 4604 & 4791 \\ 
Mathematics 2012 & 1 & 4971 & 4755 \\ 
Activity & 1 & 4975 & 5016 \\ 
Aesthetics & 1 & 4786 & 4763 \\ 
Altruism & 1 & 5082 & 4951 \\ 
Anxiety & 1 & 5023 & 4863 \\ 
Assertiveness & 1 & 5248 & 4952 \\ 
Compliance & 1 & 4707 & 4820 \\ 
Depression & 1 & 4910 & 5030 \\ 
Ideas & 1 & 4704 & 4646 \\ 
Order & 1 & 4603 & 4952 \\ 
Self-Discipline & 1 & 5080 & 4792 \\ 
\hline \\[-1.8ex] 
\end{tabular} 
\end{table}

\subsubsection{Comparing chains samplings graphically: rank plots}

The idea here is to compare the samples from the 4 chains graphically. The work \cite{vehtari2019rank} recommends that we abandon traditional Trace Plots and move on to the following Rank Plots. The idea here is: (i) we pool the samples of all chains into a single vector creating a rank (by the sampled values), (ii) compare the chains individually with the initial rank, creating a frequency histogram to evaluate whether the chains were sampled from the same distribution. If we observe uniform rank distributions across all chains, we can conclude that they were able to sample from the same distribution. In the graphs below, at least visually, we have evidence that the chains sample from the same distribution and we are more confident about the convergence of the method:

\begin{figure}[H] 
\centering 
\includegraphics[width=0.7\textwidth]{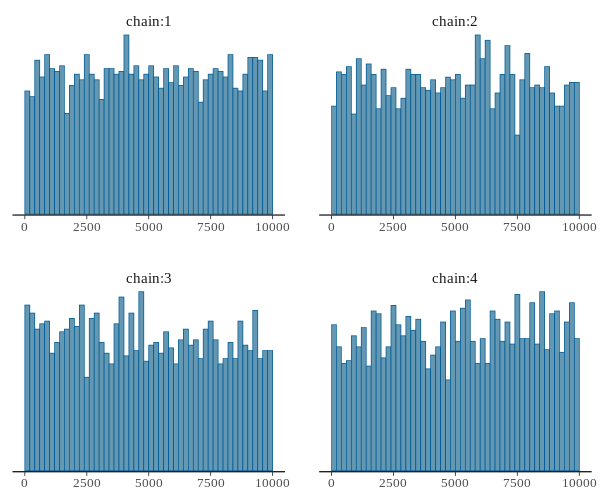} 
\caption[Caption for LOF]{Intercept}
\end{figure}

\begin{figure}[H] 
\centering 
\includegraphics[width=0.7\textwidth]{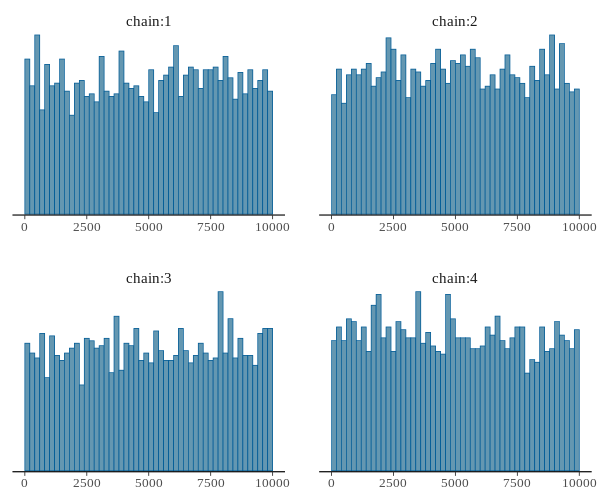} 
\caption[Caption for LOF]{grade 2012}
\end{figure}

\begin{figure}[H] 
\centering 
\includegraphics[width=0.7\textwidth]{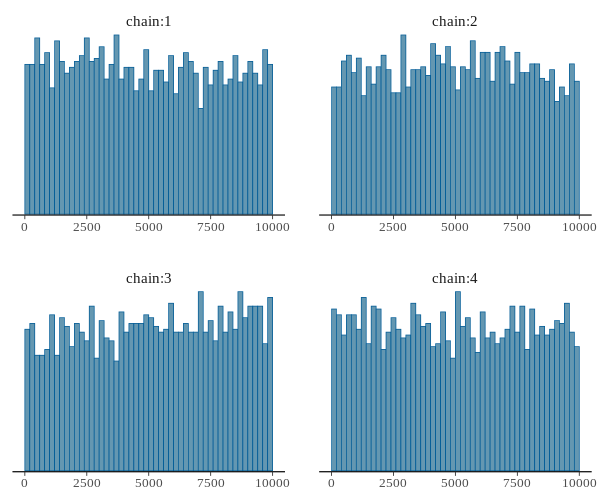} 
\caption[Caption for LOF]{year}
\end{figure}

\begin{figure}[H] 
\centering 
\includegraphics[width=0.7\textwidth]{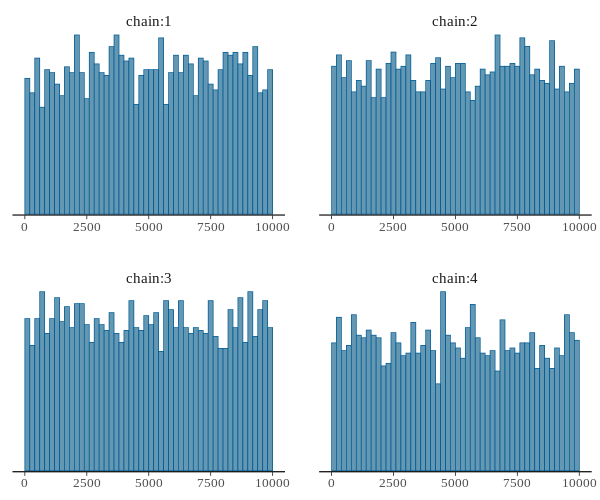} 
\caption[Caption for LOF]{semester}
\end{figure}

\begin{figure}[H] 
\centering 
\includegraphics[width=0.7\textwidth]{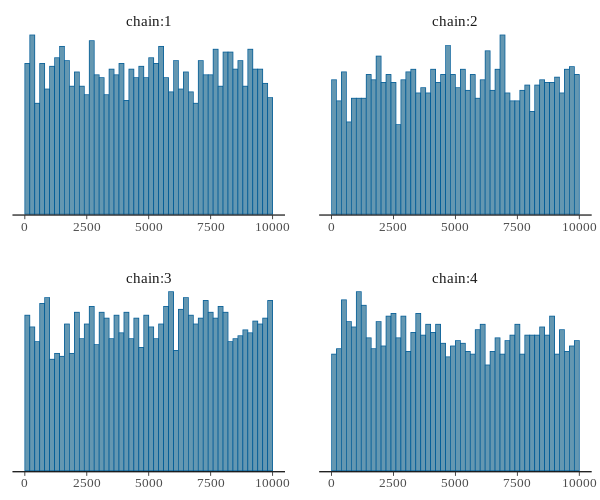} 
\caption[Caption for LOF]{school 2012 - municipal}
\end{figure}

\begin{figure}[H] 
\centering 
\includegraphics[width=0.7\textwidth]{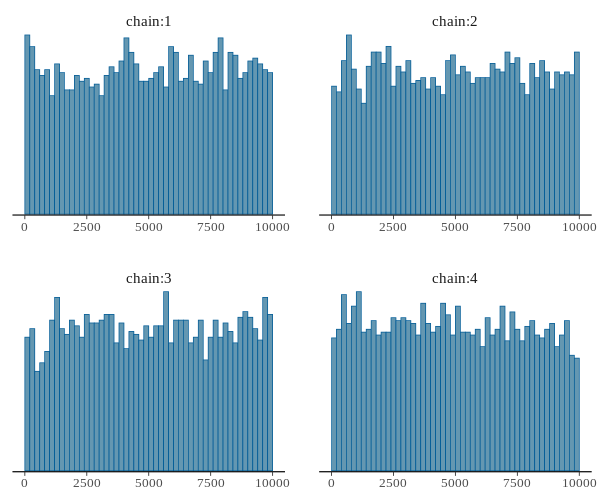} 
\caption[Caption for LOF]{school 2012 - privada}
\end{figure}

\begin{figure}[H] 
\centering 
\includegraphics[width=0.7\textwidth]{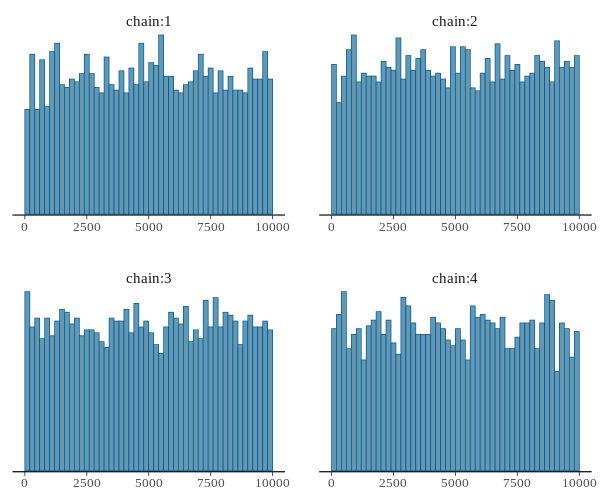} 
\caption[Caption for LOF]{white}
\end{figure}

\begin{figure}[H] 
\centering 
\includegraphics[width=0.7\textwidth]{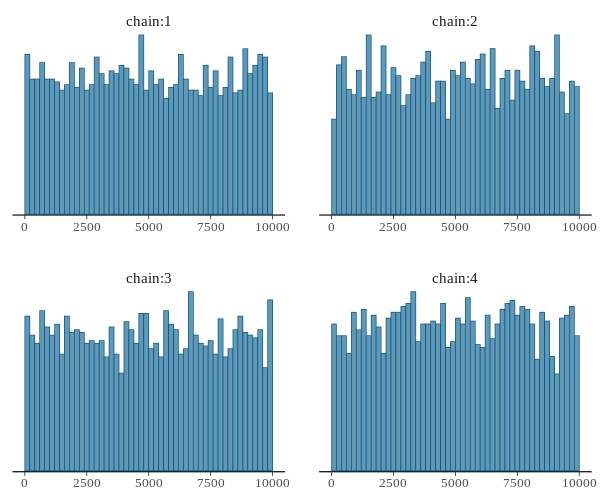} 
\caption[Caption for LOF]{male}
\end{figure}

\begin{figure}[H] 
\centering 
\includegraphics[width=0.7\textwidth]{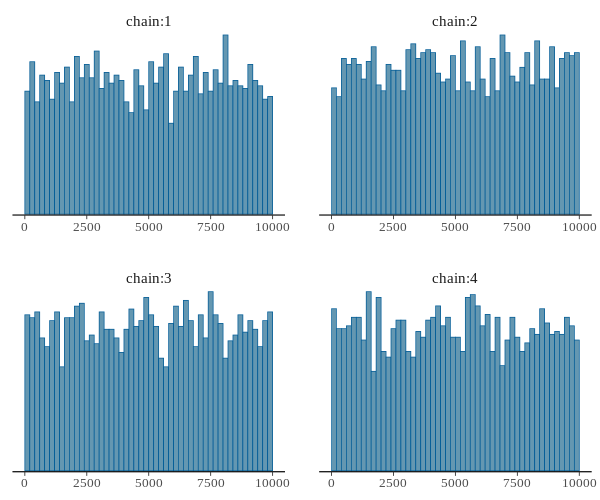} 
\caption[Caption for LOF]{failed before 2012}
\end{figure}

\begin{figure}[H] 
\centering 
\includegraphics[width=0.7\textwidth]{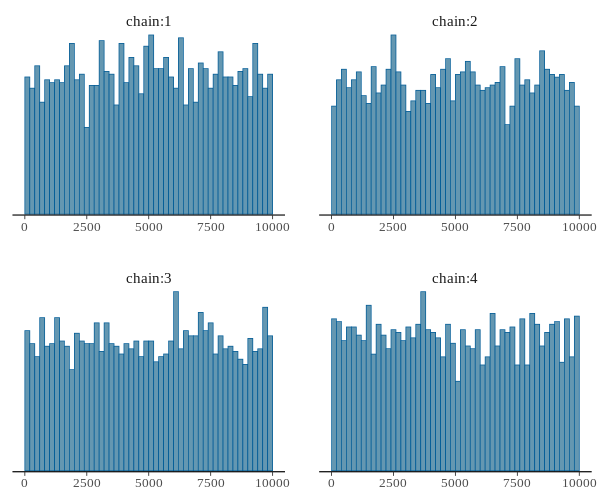} 
\caption[Caption for LOF]{mother educ - ef1}
\end{figure}

\begin{figure}[H] 
\centering 
\includegraphics[width=0.7\textwidth]{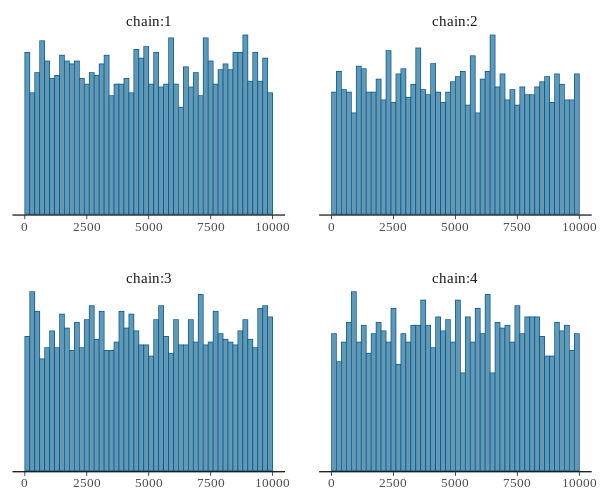} 
\caption[Caption for LOF]{mother educ - ef2}
\end{figure}

\begin{figure}[H] 
\centering 
\includegraphics[width=0.7\textwidth]{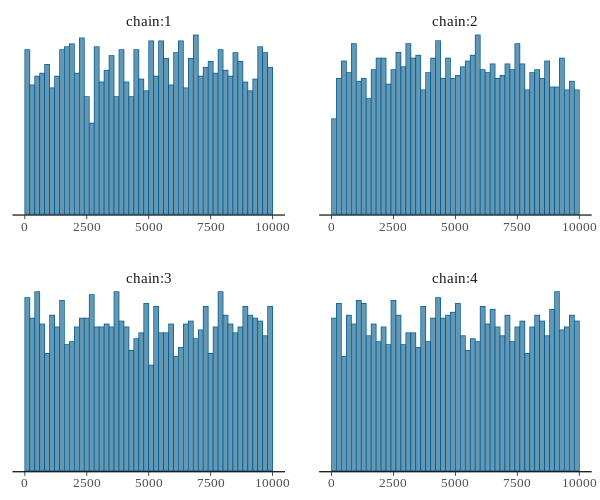} 
\caption[Caption for LOF]{mother educ - em}
\end{figure}

\begin{figure}[H] 
\centering 
\includegraphics[width=0.7\textwidth]{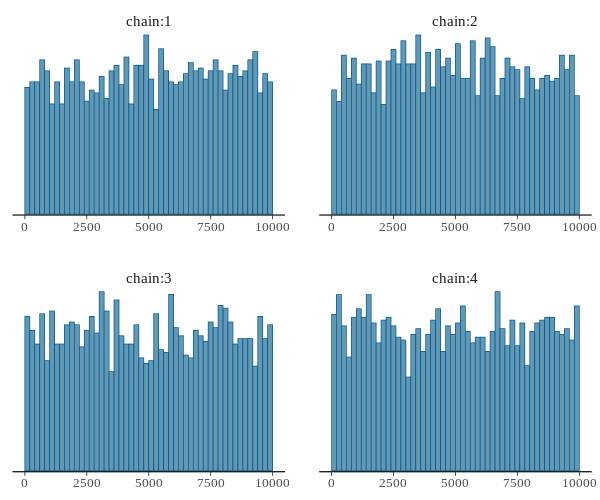} 
\caption[Caption for LOF]{mother educ - nao sabe}
\end{figure}

\begin{figure}[H] 
\centering 
\includegraphics[width=0.7\textwidth]{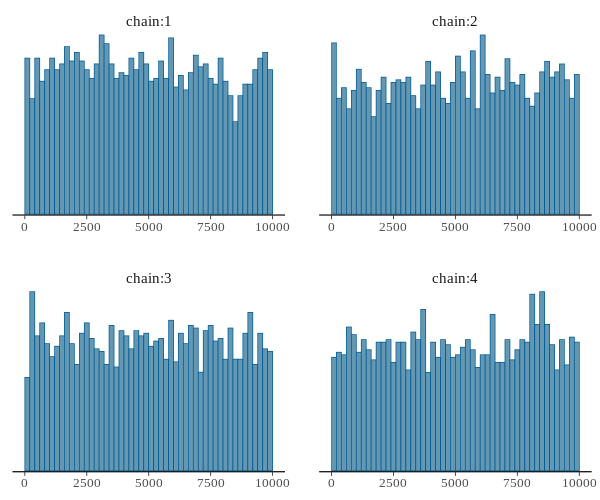} 
\caption[Caption for LOF]{mother educ - superior}
\end{figure}

\begin{figure}[H] 
\centering 
\includegraphics[width=0.7\textwidth]{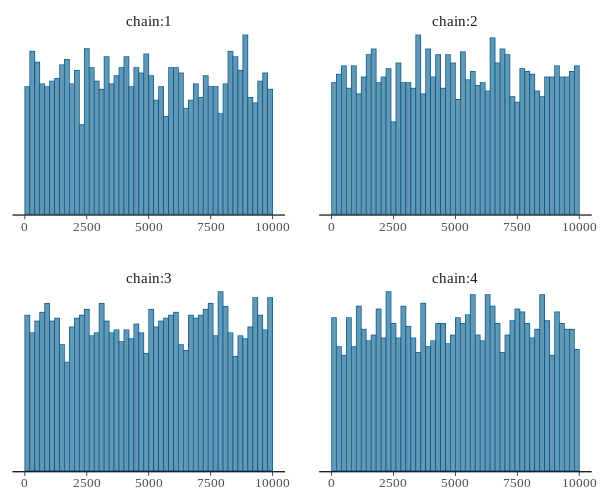} 
\caption[Caption for LOF]{pre k}
\end{figure}

\begin{figure}[H] 
\centering 
\includegraphics[width=0.7\textwidth]{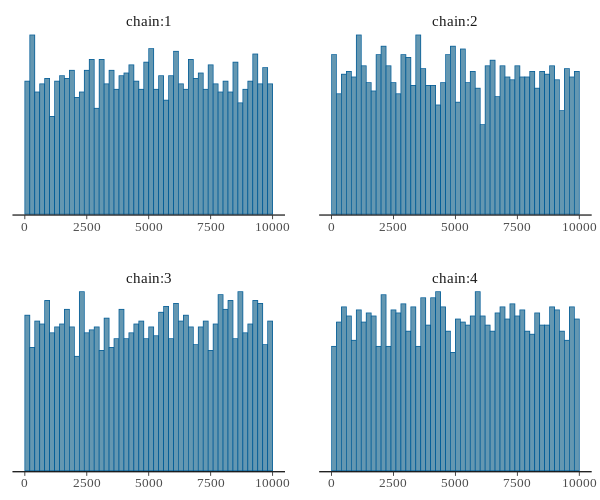} 
\caption[Caption for LOF]{kinder}
\end{figure}

\begin{figure}[H] 
\centering 
\includegraphics[width=0.7\textwidth]{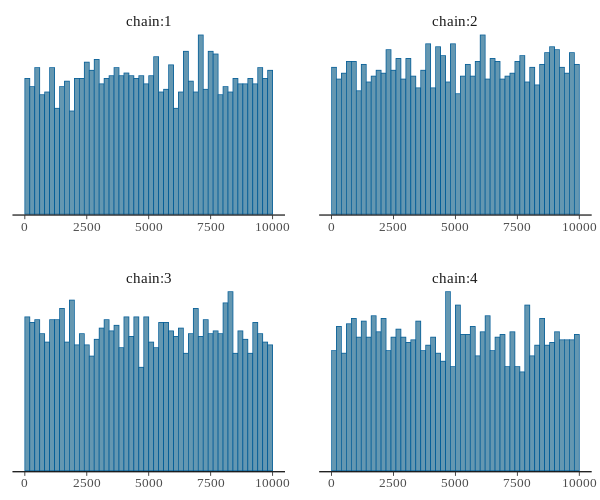} 
\caption[Caption for LOF]{lang 2012}
\end{figure}

\begin{figure}[H] 
\centering 
\includegraphics[width=0.7\textwidth]{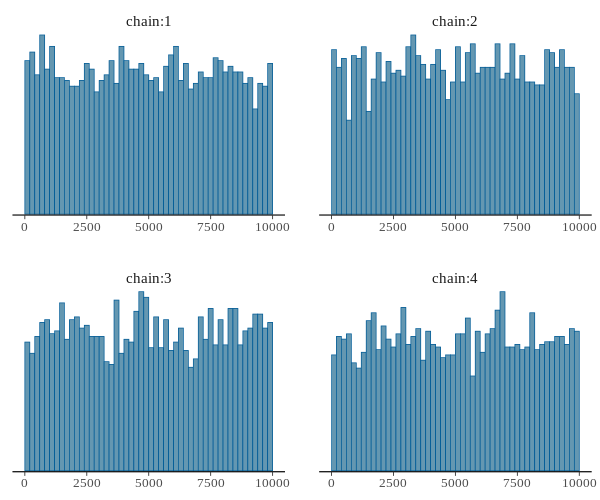} 
\caption[Caption for LOF]{math 2012}
\end{figure}

\begin{figure}[H] 
\centering 
\includegraphics[width=0.7\textwidth]{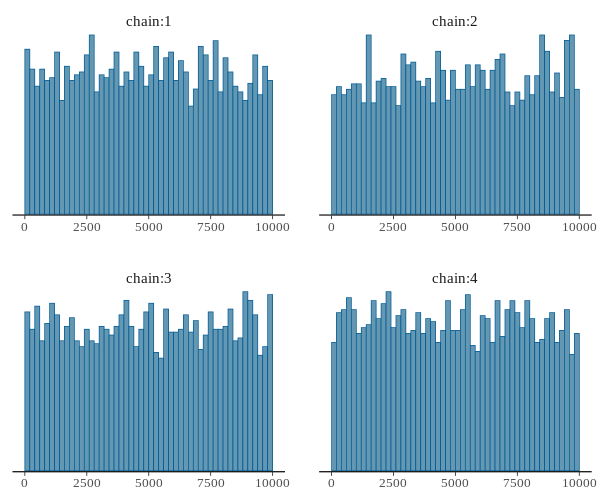} 
\caption[Caption for LOF]{act 2012}
\end{figure}

\begin{figure}[H] 
\centering 
\includegraphics[width=0.7\textwidth]{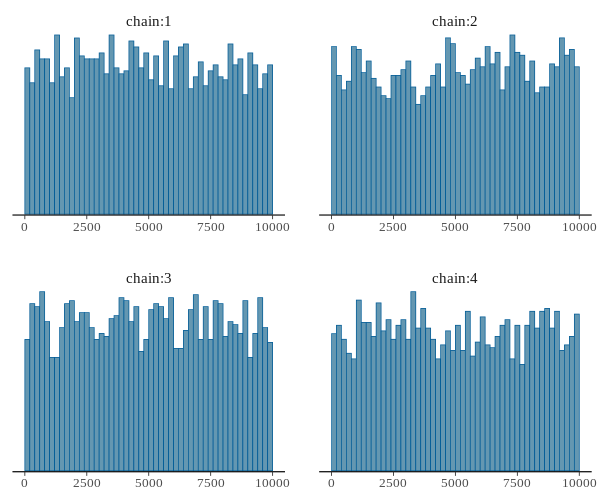} 
\caption[Caption for LOF]{aes 2012}
\end{figure}

\begin{figure}[H] 
\centering 
\includegraphics[width=0.7\textwidth]{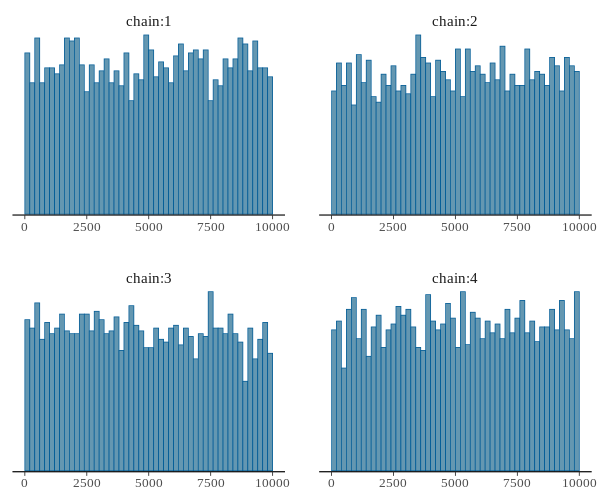} 
\caption[Caption for LOF]{alt 2012}
\end{figure}

\begin{figure}[H] 
\centering 
\includegraphics[width=0.7\textwidth]{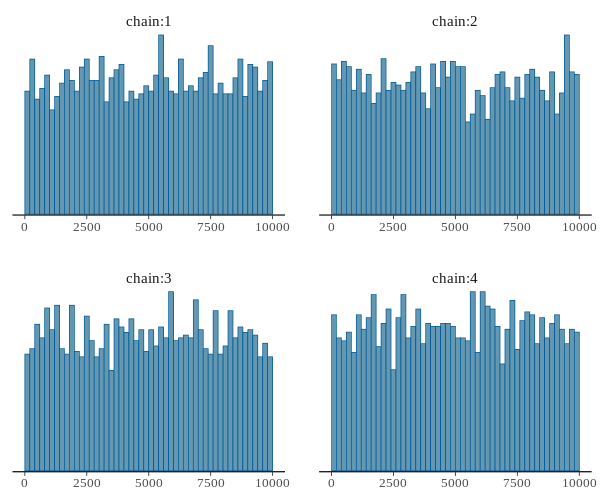} 
\caption[Caption for LOF]{anx 2012}
\end{figure}

\begin{figure}[H] 
\centering 
\includegraphics[width=0.7\textwidth]{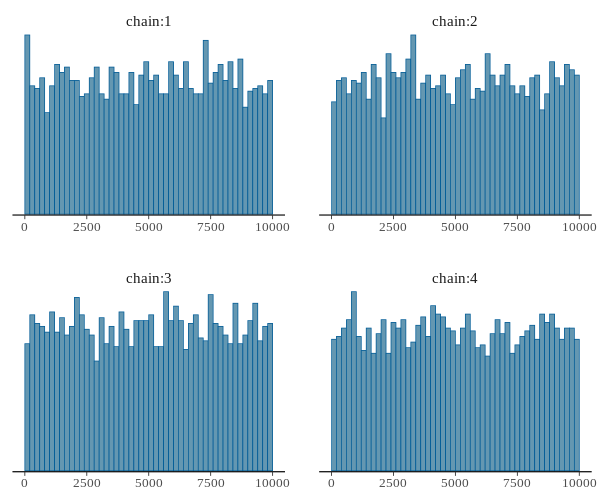} 
\caption[Caption for LOF]{ass 2012}
\end{figure}

\begin{figure}[H] 
\centering 
\includegraphics[width=0.7\textwidth]{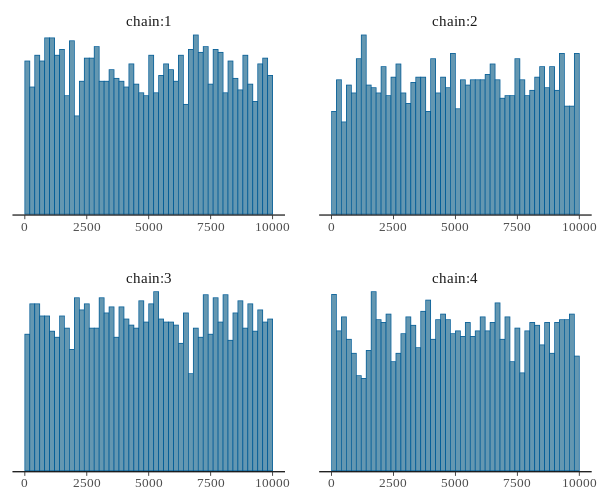} 
\caption[Caption for LOF]{cmp 2012}
\end{figure}

\begin{figure}[H] 
\centering 
\includegraphics[width=0.7\textwidth]{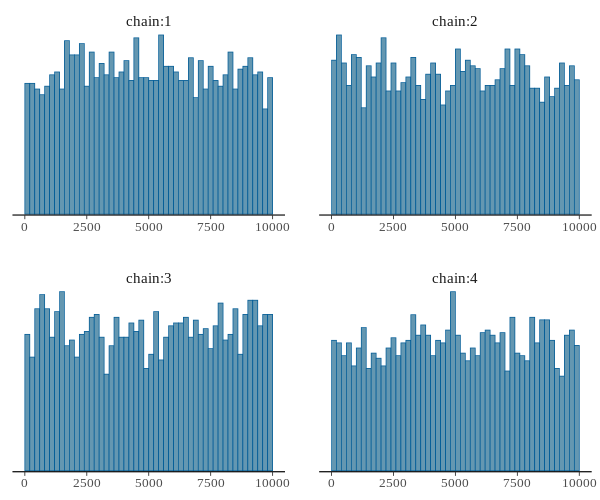} 
\caption[Caption for LOF]{dep 2012}
\end{figure}

\begin{figure}[H] 
\centering 
\includegraphics[width=0.7\textwidth]{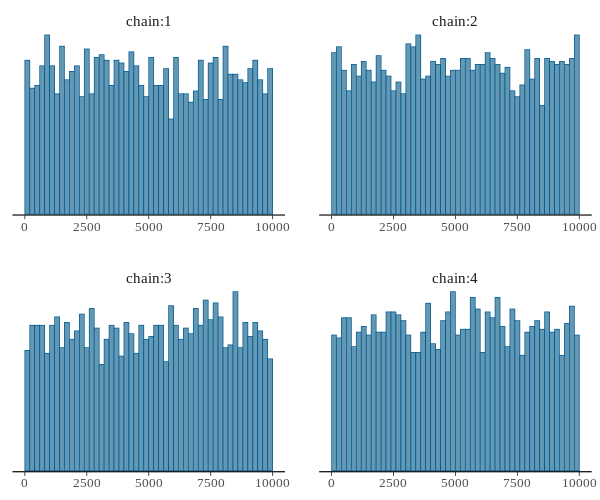} 
\caption[Caption for LOF]{ids 2012}
\end{figure}

\begin{figure}[H] 
\centering 
\includegraphics[width=0.7\textwidth]{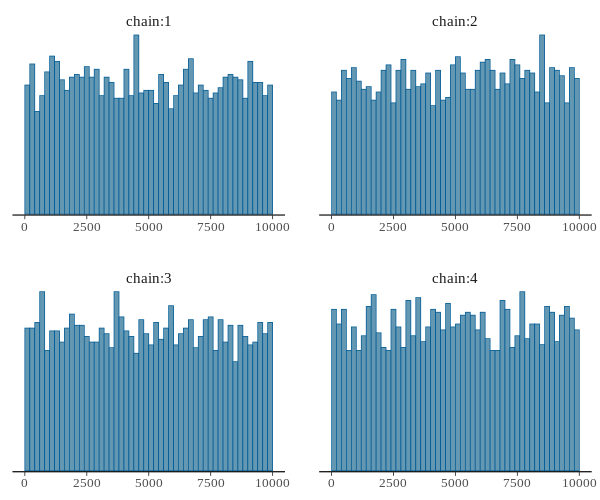} 
\caption[Caption for LOF]{ord 2012}
\end{figure}

\begin{figure}[H] 
\centering 
\includegraphics[width=0.7\textwidth]{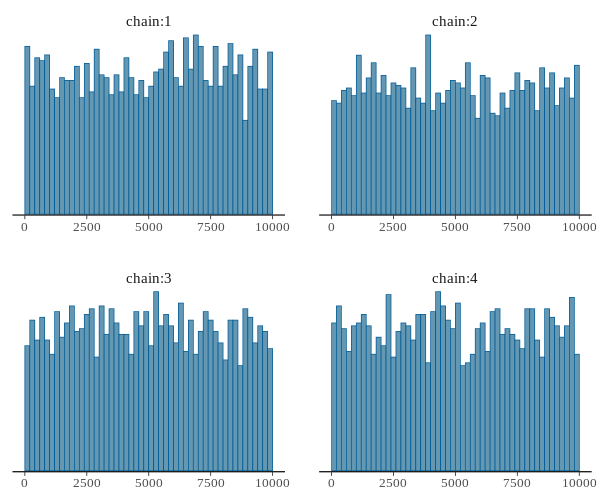} 
\caption[Caption for LOF]{sfd 2012}
\end{figure}

\subsubsection{Autocorrelation}

\begin{figure}[H] 
\centering 
\includegraphics[width=0.7\textwidth]{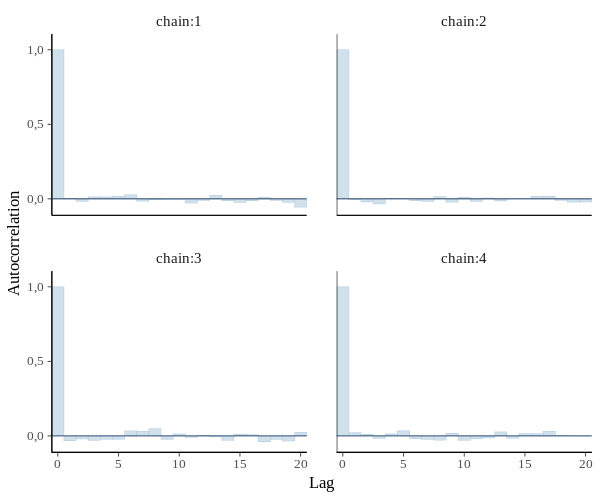} 
\caption[Caption for LOF]{Intercept}
\end{figure}

\begin{figure}[H] 
\centering 
\includegraphics[width=0.7\textwidth]{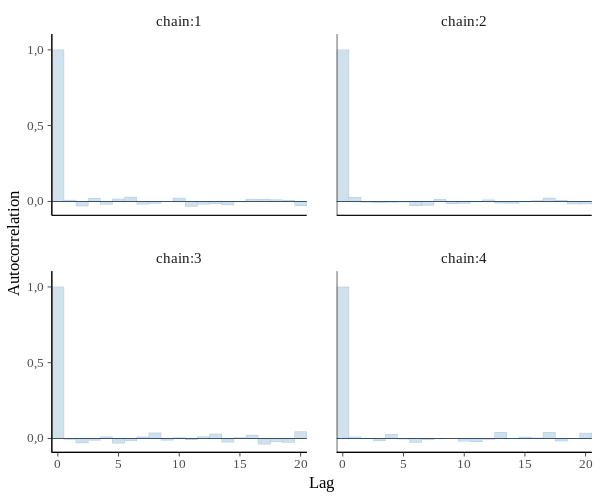} 
\caption[Caption for LOF]{grade 2012}
\end{figure}

\begin{figure}[H] 
\centering 
\includegraphics[width=0.7\textwidth]{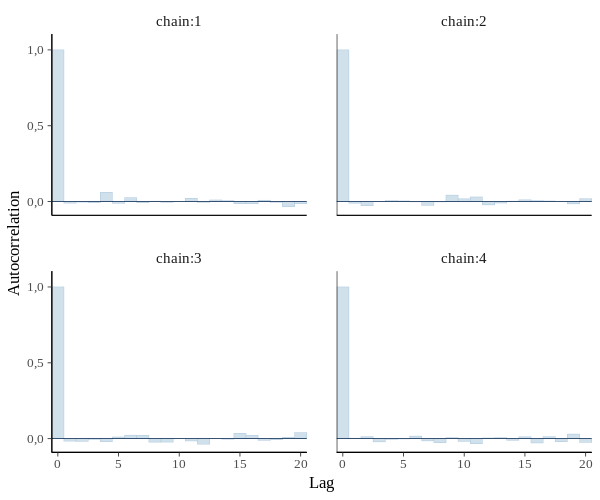} 
\caption[Caption for LOF]{year}
\end{figure}

\begin{figure}[H] 
\centering 
\includegraphics[width=0.7\textwidth]{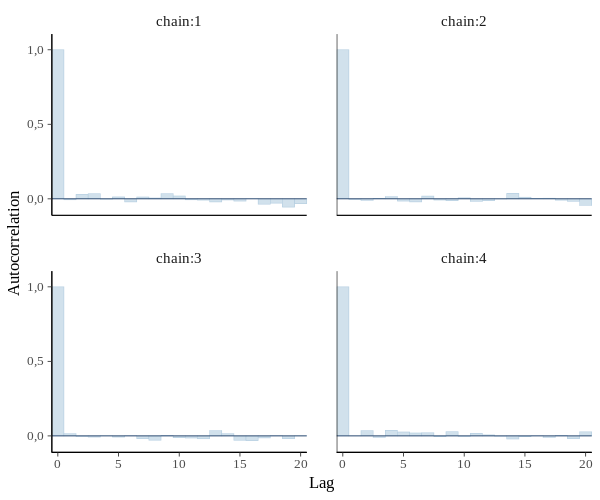} 
\caption[Caption for LOF]{semester}
\end{figure}

\begin{figure}[H] 
\centering 
\includegraphics[width=0.7\textwidth]{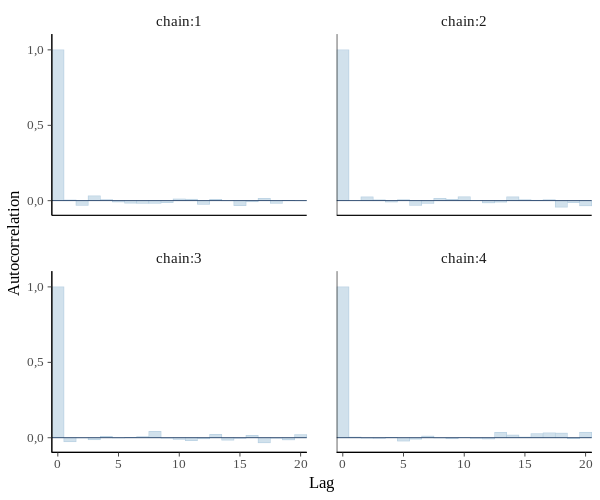} 
\caption[Caption for LOF]{school 2012 - municipal}
\end{figure}

\begin{figure}[H] 
\centering 
\includegraphics[width=0.7\textwidth]{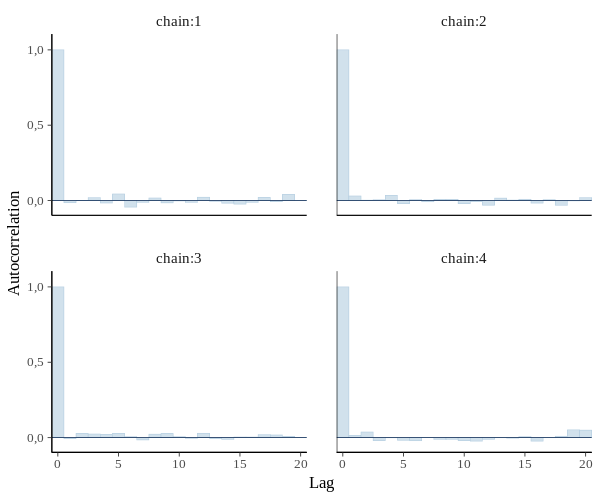} 
\caption[Caption for LOF]{school 2012 - privada}
\end{figure}

\begin{figure}[H] 
\centering 
\includegraphics[width=0.7\textwidth]{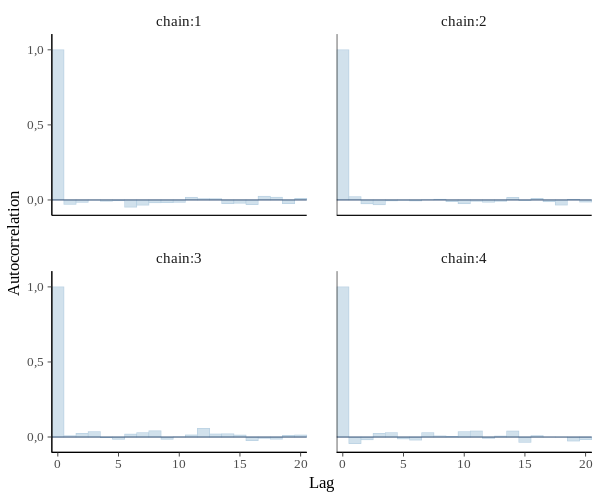} 
\caption[Caption for LOF]{white}
\end{figure}

\begin{figure}[H] 
\centering 
\includegraphics[width=0.7\textwidth]{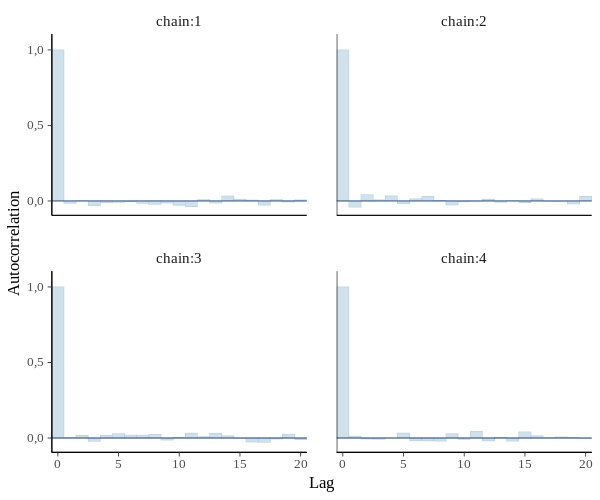} 
\caption[Caption for LOF]{male}
\end{figure}

\begin{figure}[H] 
\centering 
\includegraphics[width=0.7\textwidth]{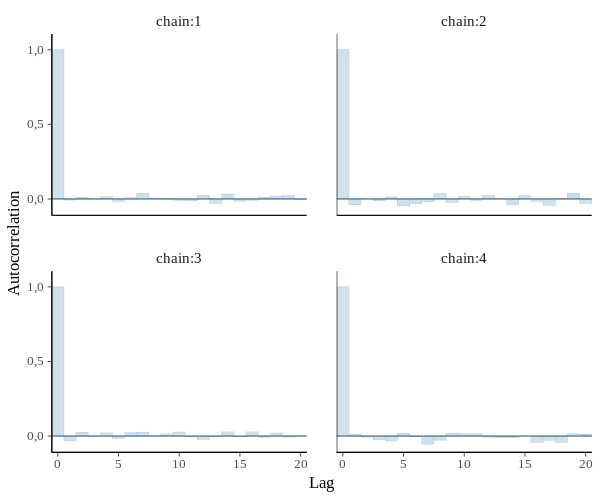} 
\caption[Caption for LOF]{failed before 2012}
\end{figure}

\begin{figure}[H] 
\centering 
\includegraphics[width=0.7\textwidth]{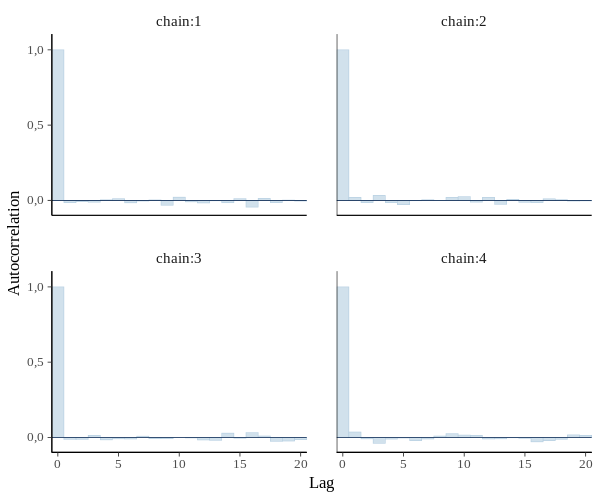} 
\caption[Caption for LOF]{mother educ - ef1}
\end{figure}

\begin{figure}[H] 
\centering 
\includegraphics[width=0.7\textwidth]{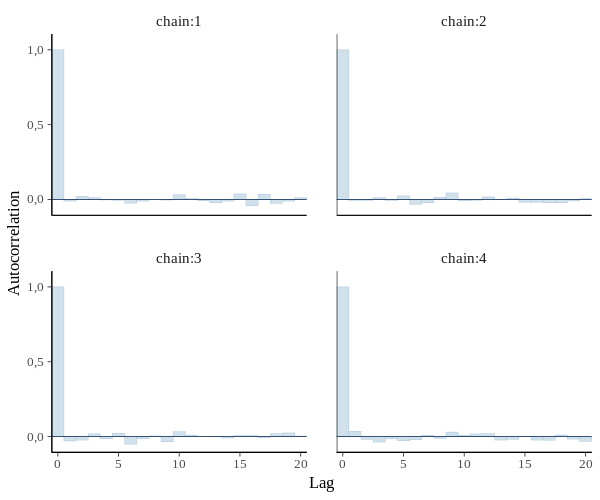} 
\caption[Caption for LOF]{mother educ - ef2}
\end{figure}

\begin{figure}[H] 
\centering 
\includegraphics[width=0.7\textwidth]{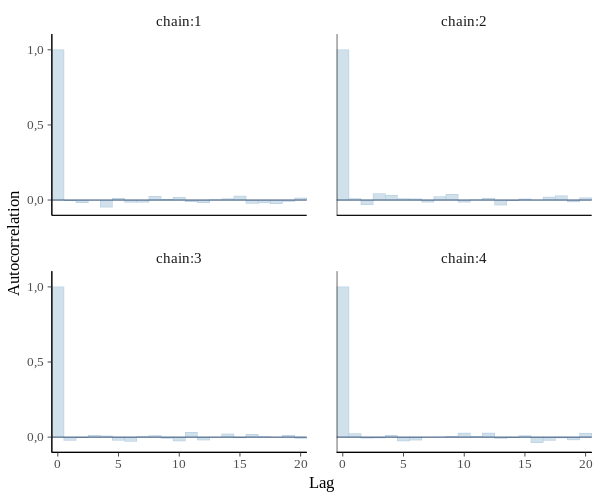} 
\caption[Caption for LOF]{mother educ - em}
\end{figure}

\begin{figure}[H] 
\centering 
\includegraphics[width=0.7\textwidth]{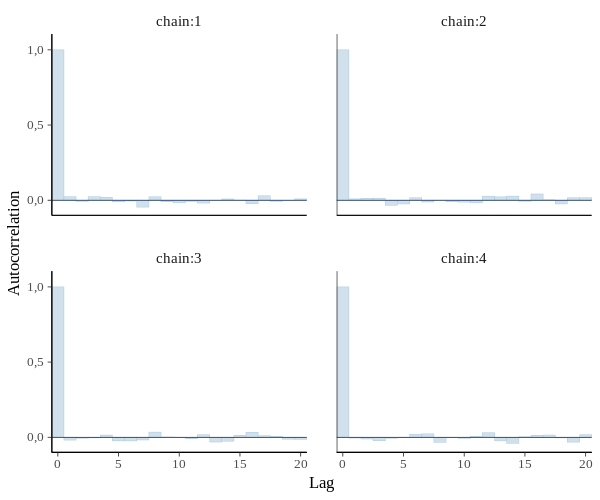} 
\caption[Caption for LOF]{mother educ - nao sabe}
\end{figure}

\begin{figure}[H] 
\centering 
\includegraphics[width=0.7\textwidth]{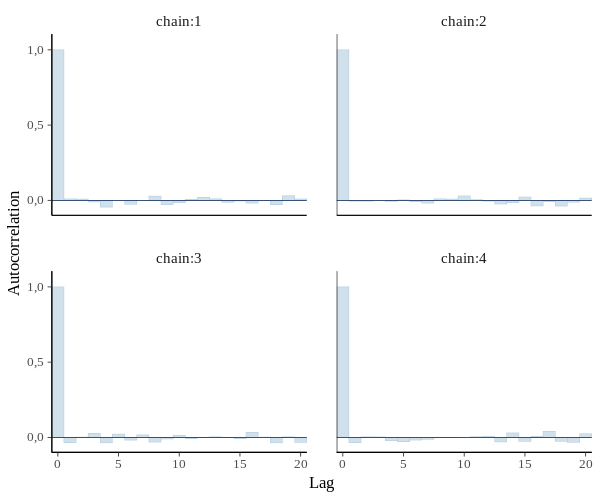} 
\caption[Caption for LOF]{mother educ - superior}
\end{figure}

\begin{figure}[H] 
\centering 
\includegraphics[width=0.7\textwidth]{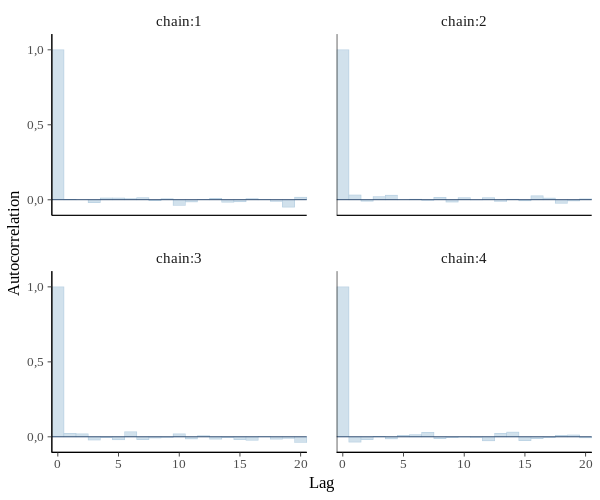} 
\caption[Caption for LOF]{pre k}
\end{figure}

\begin{figure}[H] 
\centering 
\includegraphics[width=0.7\textwidth]{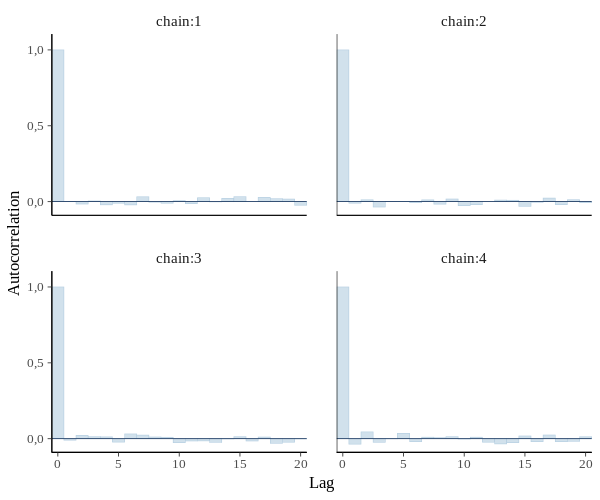} 
\caption[Caption for LOF]{kinder}
\end{figure}

\begin{figure}[H] 
\centering 
\includegraphics[width=0.7\textwidth]{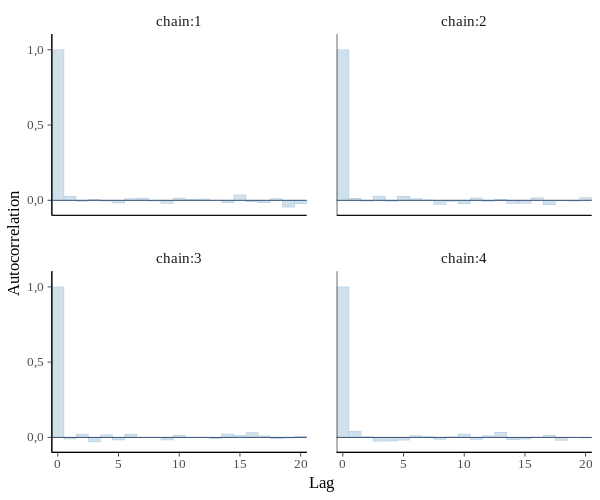} 
\caption[Caption for LOF]{lang 2012}
\end{figure}

\begin{figure}[H] 
\centering 
\includegraphics[width=0.7\textwidth]{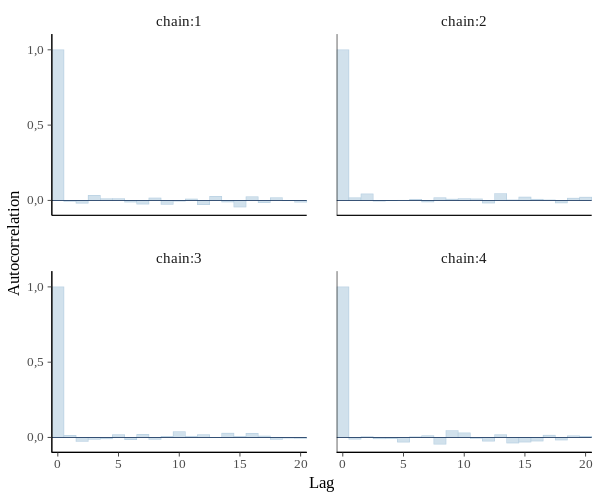} 
\caption[Caption for LOF]{math 2012}
\end{figure}

\begin{figure}[H] 
\centering 
\includegraphics[width=0.7\textwidth]{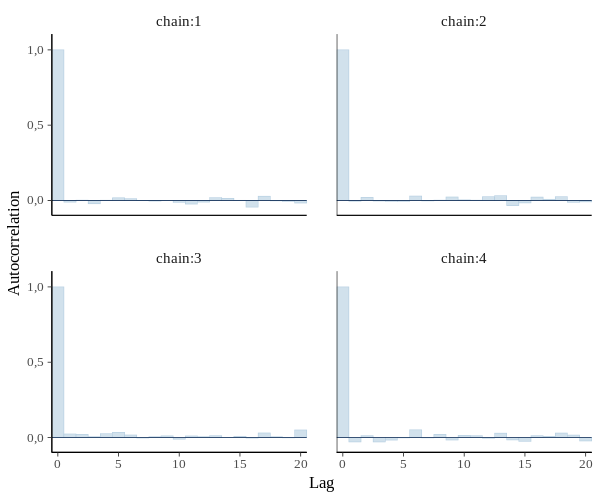} 
\caption[Caption for LOF]{act 2012}
\end{figure}

\begin{figure}[H] 
\centering 
\includegraphics[width=0.7\textwidth]{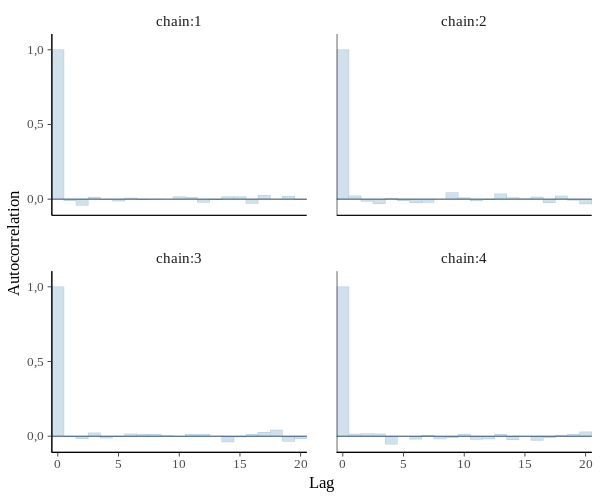} 
\caption[Caption for LOF]{aes 2012}
\end{figure}

\begin{figure}[H] 
\centering 
\includegraphics[width=0.7\textwidth]{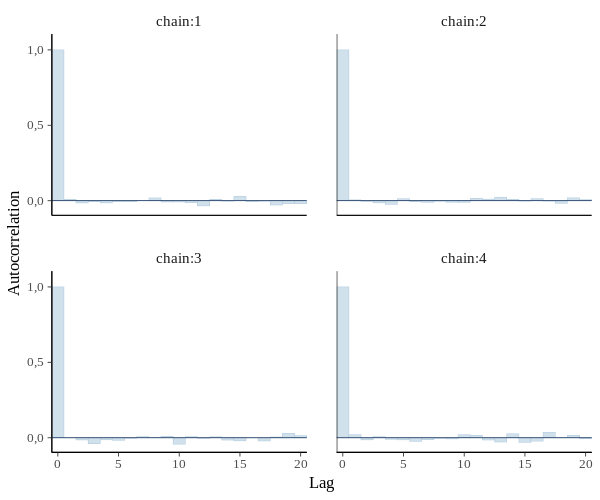} 
\caption[Caption for LOF]{alt 2012}
\end{figure}

\begin{figure}[H] 
\centering 
\includegraphics[width=0.7\textwidth]{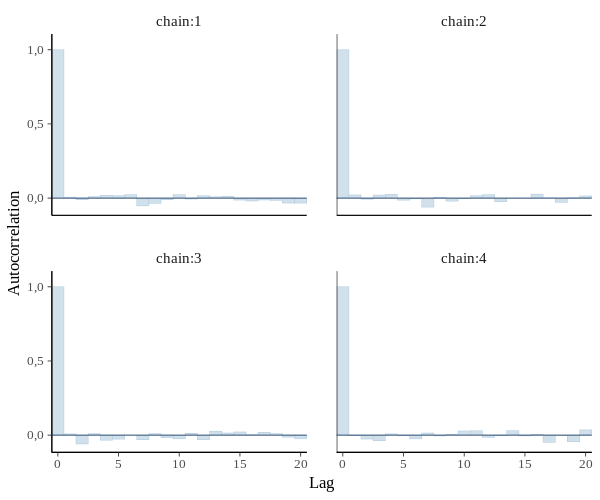} 
\caption[Caption for LOF]{anx 2012}
\end{figure}

\begin{figure}[H] 
\centering 
\includegraphics[width=0.7\textwidth]{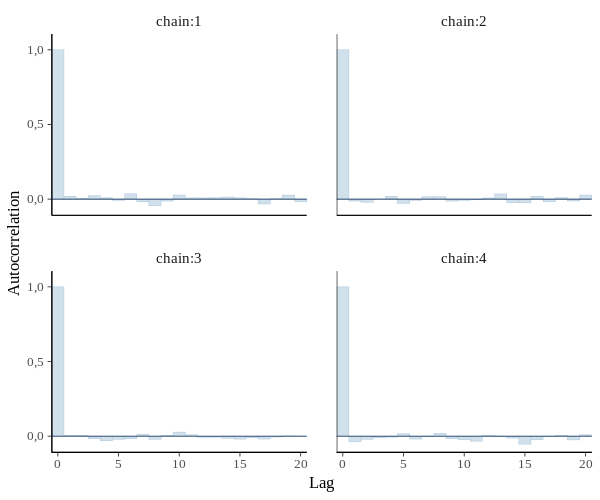} 
\caption[Caption for LOF]{ass 2012}
\end{figure}

\begin{figure}[H] 
\centering 
\includegraphics[width=0.7\textwidth]{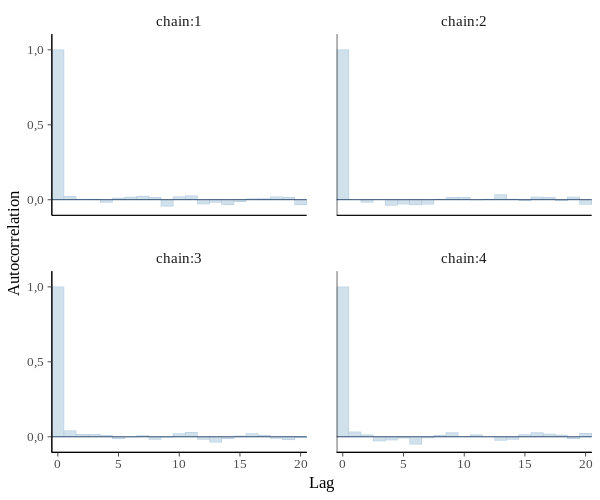} 
\caption[Caption for LOF]{cmp 2012}
\end{figure}

\begin{figure}[H] 
\centering 
\includegraphics[width=0.7\textwidth]{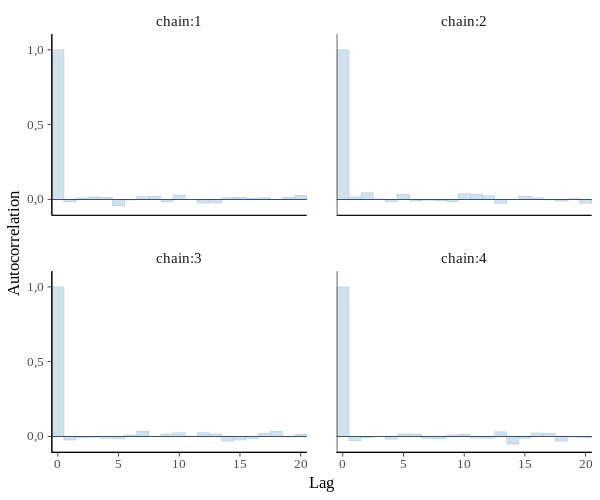} 
\caption[Caption for LOF]{dep 2012}
\end{figure}

\begin{figure}[H] 
\centering 
\includegraphics[width=0.7\textwidth]{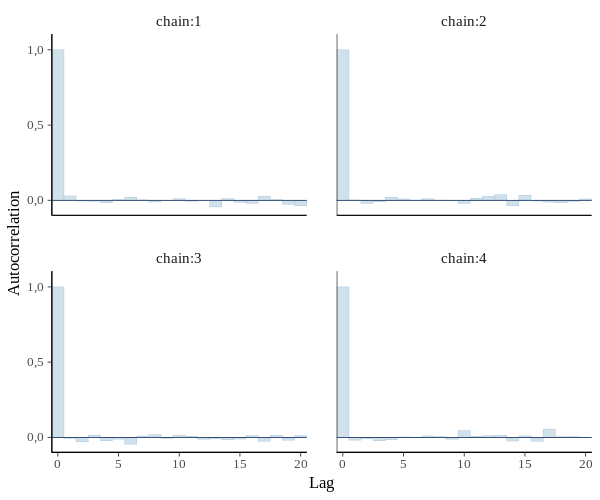} 
\caption[Caption for LOF]{ids 2012}
\end{figure}

\begin{figure}[H] 
\centering 
\includegraphics[width=0.7\textwidth]{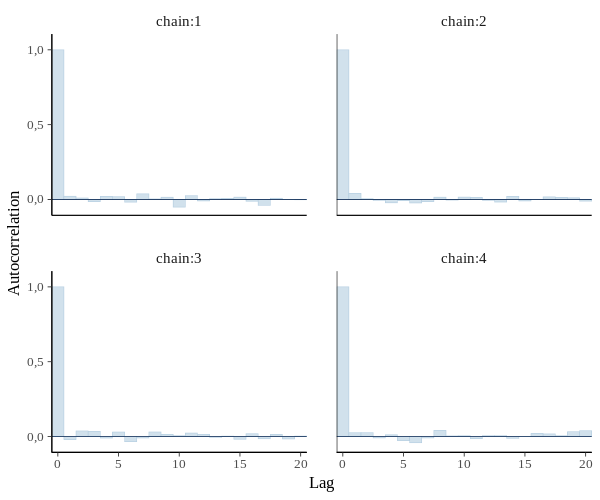} 
\caption[Caption for LOF]{ord 2012}
\end{figure}

\begin{figure}[H] 
\centering 
\includegraphics[width=0.7\textwidth]{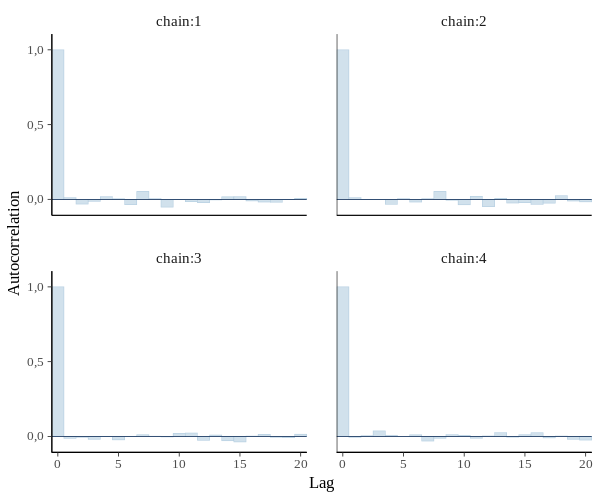} 
\caption[Caption for LOF]{sfd 2012}
\end{figure}

\end{document}